\documentclass[12pt, a4paper,reqno,]{amsart}
\usepackage{amssymb}
\usepackage{amscd}

\usepackage[all]{xy}

\theoremstyle{plain}
\newtheorem{propn}{Proposition}[section]
\newtheorem{thm}[propn]{Theorem}
\newtheorem{lem}[propn]{Lemma}
\newtheorem{cor}[propn]{Corollary}

\theoremstyle{definition}

\newtheorem{examples}[propn]{Examples}

\theoremstyle{definition}
\newtheorem*{defn}{Definition}

\newtheorem*{rem}{Remark}

\newtheorem*{exercise}{Exercise}

\newcommand{\wh}{\widehat}

\newcommand{\la}{\langle}
\newcommand{\ra}{\rangle}

\newcommand{\tu}{\textup}

\newenvironment{alist}
{

\begin{enumerate}}
{\end{enumerate}}

\newenvironment{rlist}
{

\begin{enumerate}}
{\end{enumerate}}

\newcommand{\A}{\mathcal{A}}

\newcommand{\B}{\mathcal{B}}

\newcommand{\E}{\mathcal{E}}
\newcommand{\F}{\mathcal{F}}
\newcommand{\calH}{\mathcal{H}}

\newcommand{\K}{\mathcal{K}}

\newcommand{\calL}{\mathcal{L}}

\newcommand{\N}{\mathcal{N}}

\newcommand{\Q}{\mathcal{Q}}
\newcommand{\R}{\mathcal{R}}
\newcommand{\U}{\mathcal{U}}
\newcommand{\calS}{\mathcal{S}}
\newcommand{\W}{\Gamma}

\newcommand{\NN}{\mathbb{N}}
\newcommand{\ZZ}{\mathbb{Z}}

\newcommand{\RR}{\mathbb{R}}
\newcommand{\CC}{\mathbb{C}}
\newcommand{\PP}{\mathbb{P}}
\newcommand{\EE}{\mathbb{E}}
\newcommand{\TT}{\mathbb{T}}

\def\a{\alpha}
\def\b{\beta}
\def\g{\gamma}
\def\d{\delta}
\def\e{\varepsilon}

\def\th{\vartheta}

\def\l{\lambda}

\def\r{\rho}
\def\s{\sigma}

\def\ph{\varphi}
\def\o{\omega}
\def\G{\Gamma}
\def\Si{\Sigma}
\def\O{\Omega}

\def\Re{{\rm Re}\,}
\def\Im{{\rm Im}\,}

\def\Exp{{\rm Exp}\,}

\def\tr{{\rm tr}\,}
\def\det{{\rm det}\,}
\def\sp{{\rm sp}}
\def\id{{\rm id}\,}

\def\del{\partial}
\def\li#1{\lim_{#1\to\infty}}

\def\Ltla{L^2\hbox{{-}}\lim_{\lambda}}

\def\szi#1{\sum_{#1=0}^\infty}

\def\from#1to#2{\bigg/_{\!\!\!\!\!#1}^{#2}}
\def\ssp#1#2:{\sum_{#1\in\sp({#2})}}

\def\tuple#1_#2{#1_1,#1_2,\ldots,#1_{#2}}
\def\tup#1_#2{#1_1,\ldots,#1_{#2}}
\def\seq#1{#1_1,#1_2,#1_3,\cdots}
\def\dd#1{{d\over{d#1}}}
\def\one{{\bf 1}}
\def\done{\cdot\one}
\def\Linf{L^\infty}
\def\implies{\Longrightarrow}

\def\Implies{\quad\Longrightarrow\quad}

\def\bar#1{{\overline{#1}}}
\def\intR{\int_{-\infty}^\infty}

\def\inp#1#2{\langle#1,#2\rangle}
\def\inpph#1#2{\langle#1,#2\rangle_\varphi}
\def\set#1#2{\bigl\{\>#1\>\big|\>#2\>\bigr\}}
\def\vN#1#2{{\rm vN}\bigl\{\>#1\>\big|\>#2\>\bigr\}}

\def\norm#1{\left\|\,#1\,\right\|}
\def\norms#1{\left\|\,#1\,\right\|^2}
\def\normph#1{\|#1\|_{\varphi}}

\def\ten{\otimes}
\def\sten{\otimes_{{\rm s}}}

\def\half{{\textstyle{1\over2}}}
\def\hf{{1\over2}}

\def\quarter{{\textstyle{1\over4}}}
\def\sixth{{\textstyle{1\over6}}}

\def\jtE_#1{j_{t_{#1}}(E_{#1})}

\def\St{{\widetilde \Sigma}}

\def\Ah{\widehat\A}

\def\Th{\widehat T}

\def\phh{\widehat\varphi}
\def\Aph{(\A,\ph)}
\def\Aphh{(\Ah,\phh)}
\def\gaus#1#2{e^{-{1\over2}\left(#1^2+#2^2\right)}}

\numberwithin{equation}{section}
\begin{document}

\title[]{Quantum Probability \\
applied to the Damped Harmonic Oscillator}
\author[]{Hans Maassen}
\address{Department of Mathematics, University of Nijmegen, Toernooiveld 1,
6525 ED Nijmegen, the Netherlands.}
\email{maassen@sci.kun.nl}


\begin{abstract}
In this introductory course we sketch the framework of quantum probability
in order to discuss open quantum systems,
in particular the damped harmonic oscillator.
\end{abstract}

\maketitle

\tableofcontents

\section{The framework of quantum probability}\label{one}

Noncommutative probability theory (or `quantum probability')
generalises Kolmogorov's classical probability theory
in a way that allows the inclusion of quantum mechanical models.
For a discussion of its motivation we refer to \cite{[KuM]} in this series.
Basic sources on quantum probability
outside the present series are  \cite{[Bia]},
\cite{[Dav]}, \cite{[Hol]}, \cite{[Gud]}, \cite{[Mac]},
\cite{[Mey]}, \cite{[vNe]},  \cite{[Par]}, \cite{[Var]}.
An independent introduction is given here.

\subsection{Making probability noncommutative}

In the last two decades a succesful strategy has become popular
in mathematics:
the generalisation of classical mathematical structures by noncommutative
algebraic constructions.
The most widely known example where this strategy was applied
is doubtlessly the noncommutative version of geometry,
as explained in the imaginative book of A. Connes (\cite{[Con]}).
There the classical structures of a topological space and of a
differential
manifold are the pillars on which the K-theory of C*-algebras and
a variety of cohomological algebras are built.
Another application is the field of `quantum groups',
where the classical structure of a Lie group leads into new areas in
the theory of Hopf algebras.

However, the oldest case by far
is von Neumann's and Segal's `noncommutative integration theory',
which has developed into noncommutative measure theory and probability theory.

\smallskip\noindent
The general strategy consists of the following three steps.

\begin{enumerate}
\item[(1)]
Encode the information contained in the classical structure into an appropriate
algebra of functions on it.

\item[(2)]
Characterise the resulting algebra axiomatically.
One of the axioms will be commutativity.

\item[(3)]
Drop the commutativity axiom.
\end{enumerate}

\subsubsection{Classical probability}

Let us apply this strategy to the structure of a probability space.

We remind the reader that a probability space is a triple $(\O,\Si,\PP)$,
where $\O$ is a set, $\Si$ is a $\s$-algebra of subsets of $\O$,
containing $\O$ itself,
and $\PP$ is a $\s$-additive function $\Si\to[0,1]$ with the property
that $\PP(\O)=1$ (\cite{[Kol]}).

In applications $\O$ is interpreted as the set of all possible outcomes of
a certain stochastic experiment.
$\Si$ consists of `events',
statements about the outcome of the experiment that can be tested by
observation.
When $E$ is such an event,
then $\PP(E)$ is the probability that $E$ will occur.

\subsubsection{Applying the strategy}

\textbf{Step 1}.
We choose to consider the algebra $L^\infty(\O,\Si,\PP)$,
consisting of all bounded measurable functions $f\colon\O\to\CC$,
where two such functions $f$ and $g$ are identified if $f-g$ vanishes
$\PP$-almost everywhere.
On this algebra we consider the linear functional $\ph$ given by
   $$\ph(f):=\int_\O f\,d\PP\;.$$

\smallskip\noindent
We have chosen measurable functions because we want to encode
a measurable structure.
(Had we been interested in the topological structure,
we would have chosen an algebra of continuous functions.)

Bounded functions are an appropriate choice since they
allow for unlimited multiplication of elements,
while keeping $\ph$ well-defined.

The identification of functions which are almost everywhere equal is
a technical simplification, standard in integration theory.

\smallskip\noindent
Now we must check whether all the relevant
information in $(\O,\Si,\PP)$ has been faithfully encoded.
Clearly, the triple $(\O,\Si,\PP)$ determines $L^\infty(\O,\Si,\PP)$
uniquely.
In the converse direction,
we recover a $\s$-algebra $\St$ by putting
   $$\St:=\set{p\in L^\infty(\O,\Si,\PP)}{p=p^2=p^*}\;,$$
which, however, is not isomorphic to $\Si$,
since we have identified functions that are equal almost everywhere.
In fact, $\St$ is the {\it measure algebra},
that is the quotient of $\Si$ by the equivalence
   $$S\sim T\hbox{, meaning:}\quad
     \PP\bigl((S\setminus T)\cup(T\setminus S)\bigr)=0\;.$$
This simplification is a gain rather than a loss.

Finally the probability measure $\PP$ is regained by putting
   $$\widetilde\PP:\St\to[0,1]:p\mapsto\ph(p)\;.$$

\textbf{Step 2}.
$\Linf(\O,\Si,\PP)$ is characterised as a commutative von Neumann
algebra. A few definitions should now be given.

Let $\calH$ be a Hilbert space and let $\seq A$ be a sequence of bounded
operators on $\calH$.
This sequence is said to converge to a bounded operator $A$
{\it in the strong operator topology} if for all $\psi\in\calH$:
   $$\li n\norm{A_n\psi-A\psi}=0\;.$$
It {\it increases to $A$} if, moreover,
$A_j\le A_{j+1}$ in the sense that $A_{j+1}-A_j$ is a positive operator.

A {\it von Neumann algebra} $\A$ is an algebra of bounded
operators on some
Hilbert space $\calH$ which is closed in the strong operator topology.
We shall always assume that $\A$ contains the identity operator $\one$,
and we only consider separable Hilbert spaces.

A {\it state} on a von Neumann algebra $\A$ is a linear functional
$\ph:\A\to\CC$ mapping $\one$ to the number 1,
and which is {\it positive} in the sense that $\ph(A^*A)\ge0$ for
all $A\in\A$.
If $\ph(A^*A)=0$ only for $A=0$, then $\ph$ is called {\it faithful}.
If $\lim_\lambda \ph(A_\lambda)=\ph(A)$ for every net
$(A_\lambda)$ of positive operators
increasing to $A$,
then $\ph$ is called {\it normal}.

\smallskip\noindent
Finally, for a bounded function $f$ on a probability space $(\O,\Si,\PP)$
we denote by $M_f$ the operator of multiplication by $f$ on
the Hilbert space $L^2(\O,\Si,\PP)$.

\begin{propn} \label{P 1.1}
Let $(\O,\Si,\PP)$ be a probability space.
Then the algebra
   $$\A:=\set{M_f}{f\in\Linf(\O,\Si,\PP)}$$
is a \tu{(}commutative\tu{)} von Neumann algebra of operators on $L^2(\O,\Si,\PP)$,
and the map $\ph:M_f\mapsto\int fd\PP$ is a faithful normal state on $\A$.
Conversely, every {\rm commutative} von Neumann algebra $\A$ with a faithful
normal state $\ph$ is of the above form for some classical probability space.
\end{propn}

\begin{proof}
We only prove the first part of the theorem.
The point is to show that $\A$ is strongly closed.
So let $(f_\lambda)$ be a net of $\Linf$-functions such that $M_{f_\lambda}$ tends
strongly to some bounded operator $X$ on $\calH:=L^2(\O,\Si,\PP)$, that is
for all $\psi\in\calH$ we have
   $$\Ltla f_\lambda \psi=X\psi\;.$$
Without loss of generality we may assume that $\norm{X}=1$.
We must show that $X=M_f$ for some $f\in\Linf$.
Put $f:=X1$.
Then for all $g\in\Linf$ we have
   $$Xg=\Ltla f_\lambda g=\Ltla M_gf_\lambda =M_g\left(\Ltla f_\lambda \cdot 1\right)
       =M_gX1=M_g f=fg\;.$$
Now let the event $E_\e$ for $\e>0$ be defined by
   $$E_\e:=\set{\o\in\O}{\;|f(\o)|^2\ge1+\e}\;.$$
Then, since $\norm{X}\le1$,
   $$\PP(E_\e)=\norms{1_{E_\e}}\ge\norms{X1_{E_\e}}=\norms{f1_{E_\e}}
              =\int_{E_\e}|f|^2d\PP\;\ge(1+\e)\PP(E_\e)\;,$$
and it follows that $\PP(E_\e)=0$.
Since this holds for all $\e>0$,
we have $|f|\le1$ almost everywhere with respect to $\PP$.
So $f\in\Linf(\O,\Si,\PP)$.
Finally, since the operators $X$ and $M_f$ are both bounded and coincide
on the dense subspace $\Linf$ of $\calH$, they are equal.\end{proof}

\textbf{Step 3}.
We now drop the commutativity requirement to arrive at the following
definition.

\begin{defn}
By a {\it noncommutative probability space} we mean a pair $(\A,\ph)$,
where $\A$ is a von Neumann algebra of operators on some Hilbert space $\calH$,
and $\ph$ is a normal state on $\A$.
If $\ph$ is faithful, the probability space is called {\it non-degenerate}.
\end{defn}

\subsection{Events and random variables} \label{sub 1.2}

Let us carry some important concepts of probability theory
over from the structure $(\O,\Si,\PP)$ to the generalised probability space
$(\A,\ph)$.
(Comparable discussions are found in \cite{[Mac]},
\cite{[KuM]}, \cite{[Par]}, \cite{[Mey]}.)

\smallskip\noindent
Classically an event is an element $S$ of $\Si$.
In Step 1 of the preceeding section this is replaced by the projection
$1_S$ in $\Linf(\O,\Si,\PP)$.
In Step 3 the concept of an event is generalised to that of an arbitrary
orthogonal projection in $\A$, that is an element satisfying $E^2=E=E^*$.
The state $\ph$ associates to this event the probability $\ph(E)$.
The operator 0 is the impossible event, and $\one$ is the sure event.
Two events $E$ and $F$ are called {\it compatible} if $EF$ is also
an event, equivalently $E$ and $F$ commute:
   $$
   EF=(EF)^* \Leftrightarrow  EF(=(EF)^*) =FE \Rightarrow
   (EF)^2 = E^2F^2 =EF\;.
   $$
If this is the case, then the event $EF$ stands for the occurrence of
both $E$ and $F$ and the event $E\vee F:=E+F-EF$ for the occurrence
of either $E$ or $F$ or both.
If $EF=0$, then the occurrences of $E$ and $F$ exclude each other.
So mutually exclusive events are described by orthogonal subspaces of $\calH$.

\smallskip\noindent
A classical random variable on a probability space $(\O,\Si,\PP)$
is a measurable function $X$ from $\O$ to some other measure space
$(\O',\Si')$.
Such a function induces an embedding of $\Si'$ into $\Si$ given by
           $$J_X:S\mapsto X^{-1}(S),$$
containing the same information as $X$ itself.
The probability distribution of $X$ is given by
           $$\PP_X:=\PP\circ J_X:S\mapsto\PP(X^{-1}(S))\;.$$
Let us see what our program does with this structure.
In Step 1 the embedding $J_X$ is replaced by the mapping
   $$j_X:\Linf(\O',\Si',\PP_X)\to\Linf(\O,\Si,\PP):f\mapsto f\circ X\;,$$
the natural extension of the map $1_S\mapsto E(S):=1_{X^{-1}(S)}$
to an (injective) *-homomorph-ism $\Linf(\O',\Si',\PP_X)\to\Linf(\O,\Si,\PP)$.
Still the projection $E(S)$ stands for the classical event
$X^{-1}(S)$ that the random variable $X$ takes a value in $S\in\Si'$.

In Step 3 this is now generalised to the following notion.

\begin{defn}
By a {\it generalised random variable} on a noncommutative
probability space
$(\A,\ph)$ we mean a *-homomorphism from some other von Neumann algebra
$\B$ into $\A$ mapping $\one_\B$ to $\one_\A$.
The {\it probability distribution} of $j$ is the state $\psi:=\ph\circ j$
on $\B$.
\end{defn}

We denote this state of affairs briefly by
   $$j:(\B,\psi)\to(\A,\ph)\;.$$
If $\B$ is commutative, say $(\B,\psi)=\Linf(\O',\Si',\PP')$,
then the random variable $j$ is said to {\it take values in} $\O'$,
and $j$ can be written
   $$j(f)=\int_{\O'}f(\l)E(d\l)\;,$$
where $E$ denotes the projection-valued measure given by
   $$E(S):=j(1_S),\quad(S\in\Si').$$
In the particular case that $\O'=\RR$,
$j$ determines a unique self-adjoint operator on the representation space
$\calH$ of $\A$:

\begin{thm}[Spectral Theorem, von Neumann]
There is a one-to-one correspondence between self-adjoint operators
$A$ on a Hilbert space $\calH$ and projection-valued measures
$E:\Si(\RR)\to\B(\calH)$ such that
   $$A=\int_\RR\l E(d\l)\;.$$

\end{thm}

When $E(S)\in\A$ for all $S$ in the Borel $\s$-algebra $\Si(\RR)$,
then $A$ is said to be {\it affiliated} to $\A$.
Moreover:

\begin{thm}[Stone's Theorem] \label{stone}
There is a one-to-one correspondence between strongly continuous
unitary representations $t\mapsto U_t$ of the abelian group $\RR$
into $\A$ and self-adjoint operators $A$ affiliated to $\A$ such that
   $$U_t=e^{itA}\;.$$
\end{thm}

Here the right hand side is to be read as the strongly convergent integral
   $$e^{itA}:=\int_\RR e^{it\l}E(d\l)\;,$$
where $E$ is given by the spectral theorem.
If we put $e_t:\RR\to\CC:x\mapsto e^{itx}$, then the connection with $j$
can be written
   $$j(e_t)=e^{itA}\;.$$

\smallskip\noindent
So altogether we can characterise a real-valued random variable
or {\it observable} in any one of four ways:

\begin{enumerate}
\item[(1)]
by a self-adjoint operator $A$ affiliated to $\mathcal{A}$;
\item[(2)]
by a projection-valued measure $E$ in $\A$;
\item[(3)]
by a normal injective *-homomorphism $j:\Linf(\RR,\Si(\RR),\PP)\to\A$; and
\item[(4)]
by a one-parameter unitary group $(U_t)_{t \in \RR}$ in $\A$.

\end{enumerate}

\subsection{Interpretation of quantum probability}\label{sub 1.3}

It is a surprising fact that nature --- at least on small scales ---
appears to be governed by noncommutative probability.

Quantum probability describes manipulations performed on physical systems by
certain mappings between generalised probability spaces called
{\it operations}.
(The same has been said about classical probability see, e.g. \cite{[vKa]}.)
These mappings will be treated in some detail in Section \ref{three}.
The generalised random variable which we just saw is such a mapping.
It represents the operation of restricting attention to a subsystem.
Another such mapping is the conditional expectation,
describing the immersion of a physical system into a larger one.
Yet other operations are the time evolution and the transition operator:
they represent the act of waiting for some time while the system
evolves on its own, or in interaction with something else respectively.

At the end of a chain of operations we land in some probability space
$(\A,\ph)$, and we need a way to interpret it in terms of the outcome of
the physical experiment which is being described.
The rules for interpretation are as follows.

\begin{itemize}
\item
Some of the orthogonal projections in $\A$ have an
interpretation as observable events.

\end{itemize}

In the language of Mackey (\cite{[Mac]}) events may be considered as
{\it questions}, which can be asked to the system.
It answers by saying `yes' or `no'.

\begin{itemize}

\item
The experiment can be repeated arbitrarily often.
Each time we are free to choose new questions,
that is we are allowed to adjust observation equipment without our whole
experiment becoming a different one.
(This is the main distinction with the stochastic experiments
envisaged by Kolmogorov in \cite{[Kol]}.)

\item
Compatible questions can be asked together in the same trial.

\item
Incompatible questions can be asked in different trials.

\item
Inside one single trial it is sometimes possible to ask
incompatible questions one {\it after} the other.
The order will then influence the probabilities:
if the questions $\tuple E_k$ are asked in each trial and in this order,
the asymptotic fraction of the trials in which they are all answered `yes'
is
   $$\ph(E_1E_2\cdots E_{k-1}E_kE_{k-1}\cdots E_2E_1)\;.$$

\end{itemize}

\begin{rem}
It is sometimes difficult to say where the operation ends and the observation
begins.
For example, posing the question $E$ after the question $F$ can
alternatively be viewed
as an {\it operation} $l^\infty(\{0,1\}\times\{0,1\})\to\A$,
followed by the observation of the compatible events
$\{1\}\times\{0,1\}$ and $\{0,1\}\times\{1\}$.

\end{rem}


\subsection{The quantum coin toss: `spin'}

The simplest noncommutative von Neumann algebra is $M_2$,
the algebra of all $2\times2$ matrices with complex entries.
And the simplest noncommutative probability space is $(M_2,\half\tr)$,
the `fair quantum coin toss'.

The events in this probability space are the orthogonal projections
in $M_2$: the complex $2\times2$ matrices $E$ satisfying
   $$E^2=E=E^*\;.$$
Let us see what these projections look like.
Since $E$ is self-adjoint,
it must have two real eigenvalues,
and since $E^2=E$ these must both be 0 or 1.
So we have three possibilities.

\begin{enumerate}
\item[(0)]
Both are 0; that is $E=0$, the impossible event.

\item[(1)]
One of them is 0 and the other is 1.

\item[(2)]
Both are 1; that is $E=\one$, the sure event.
\end{enumerate}

In Case (1),
$E$ is a one-dimensional projection satisfying
   $$\tr E=0+1=1 \, \text{ and } \, \det E=0\cdot1=0\;.$$
As $E^*=E$ and $\tr E=1$ we may write
   $$E=\half\begin{bmatrix}1+z&x-iy \\ x+iy&1-z \end{bmatrix}\;,
      \quad\hbox{\rm with }(x,y,z)\in\RR^3\;.$$
Then $\det E=0$ implies that
   $$\quarter((1-z^2)-(x^2+y^2))=0\Implies x^2+y^2+z^2=1\;.$$
So the one-dimensional projections in $M_2$ are parametrised by the
unit sphere $S_2$.
\smallskip

\noindent \textbf{Notation}.
For $a=(a_1,a_2,a_3)\in\RR^3$ let us write
   $$\s(a):=\begin{bmatrix}a_3&a_1-ia_2 \\ a_1+ia_2&-a_3\end{bmatrix}
           =a_1\s_1+a_2\s_2+a_3\s_3\;,$$
where $\s_1,\s_2$ and $\s_3$ are the {\it Pauli matrices}
   $$\s_1:=\begin{bmatrix}0&1\\ 1&0\end{bmatrix}\;,\quad
     \s_2:=\begin{bmatrix}0&-i\\ i&0 \end{bmatrix}\;,\quad
     \s_3:=\begin{bmatrix}1&0\\ 0&-1 \end{bmatrix}\;.$$
We note that for all $a,b\in\RR^3$ we have
   \begin{equation}  \label{onea}
   \s(a)\s(b)=\inp ab\done+i\s(a\times b)\;.
   \end{equation}

Let us write
\begin{equation}  \label{twoa}
   E(a):=\half(\one+\s(a)),\quad(\norm a=1)\;.
   \end{equation}

In the same way the possible states on $M_2$ can be calculated.
We find that
\begin{equation} \label{threea}
 \ph(A)=\tr(\r A)\text{ where }
 \r=\r(a):=\half(\one+\s(a)),\quad\norm a\le1\;.
 \end{equation}

The situation is summarised by the following proposition.

\begin{propn}\label{P 1.5}
The states on $M_2$ are parameterised by the unit ball in $\RR^3$,
as in \tu{\eqref{threea}},
and the one-dimensional projections in $M_2$ are
parametrised by the unit sphere as in \tu{\eqref{twoa}}.
The probability of the event $E(a)$ in the state $\r(b)$ is given by
   $$\tr(\r(b)E(a))=\half(1+\inp ab)\;.$$
The events $E(a)$ and $E(b)$ are compatible if and only if $a=\pm b$.
Moreover we have for all $a\in S_2$:
   $$E(a)+E(-a)=\one\;,\quad E(a)E(-a)=0\;.$$
\end{propn}

\noindent \emph{Interpretation}.
The probability distribution of the quantum coin toss or {\it `qubit'}
is given by a unit vector $b$ in three-dimensions.
For every $a$ on the unit sphere we can say with probability one that
of the two events $E(a)$ and $E(-a)$ exactly one will occur,
$E(a)$ having probability $\half(1+\inp ab)$.
We therefore have, for each direction $a$, a classical coin toss
with probability for heads equal to $\half(1+\inp ab)$.
The coin tosses in different directions are incompatible.

Particular case: the quantum fair coin is modelled by $(M_2,\half\tr)$.

The quantum coin toss is realised in nature:
the spin direction of a particle with total spin $\half\hbar$ behaves in
this way.

\subsection{Positive definite kernels} \label{sub 1.5}

In this section we introduce a useful tool for the construction of
Hilbert spaces, used heavily in quantum probability.

Let $\calS$ be a set and let $K$ be a {\it kernel} on $\calS$,
that is a function $\calS\times\calS\to\CC$.
Then $K$ is called {\it positive definite} if for all $n\in\NN$ and
all $n$-tuples $(\tup\l_n)\in\CC^n$ we have
$$
\sum_{i=1}^n\sum_{j=1}^n\bar{\l_i}\l_j K(x_i,x_j)\ge0\;.
$$

\begin{thm}[Kolmogorov's dilation theorem] \label{T 1.6}
Let $K$ be a positive definite kernel on a set $\calS$.
Then up to unitary equivalence there exists a unique Hilbert space $\calH$
and a unique embedding $V:\calS\to\calH$ such that
\begin{align}
&\forall_{x,y\in\calS}:\inp{V(x)}{V(y)}=K(x,y) \label{oned} \\
&\tu{\text{ and }} \, \bar{\bigvee V(\calS)}=\calH\;. \label{onee}
\end{align}
\end{thm}
A map $V:\calS \rightarrow \calH$ is called a (\emph{Kolmogorov})
\emph{dilation} if \eqref{oned} holds. It is called \emph{minimal}
if \eqref{onee} holds.

\begin{proof}
Consider the space $\calL$ of all functions $\calS\to\CC$ with finite support.
Then $\calL$ becomes a pre-Hilbert space if we define the (pre-)inner product
$$
\inp\l\mu:=\sum_{x\in\calS}\sum_{y\in\calS}\bar{\l(x)}K(x,y)\mu(y)\;.
$$
Dividing out the null space
$$
\N:=\set{\l\in\calL}{\inp\l\l=0}
$$
and forming the completion $\calH_K$ of $\calL/\N$,
we let $V_K:\calS\to\calH_K$ be given by
$$
V_K(x):=\d_x+\N\;.
$$
Then for all $x,y\in\calS$:
$$
\inp{V_K(x)}{V_K(y)}=\inp{\d_x+\N}{\d_y+\N}_{\calL/\N}
                    =\inp{\d_x}{\d_y}_\calL=K(x,y)\;.
$$
Now let $V:\calS\to\calH$ be a second minimal Kolmogorov dilation of $K$.
Then we define a map
$$
U_0:\calL\to\calH:\quad\l\mapsto\sum_{x\in\calS}\l(x)V(x)\;.
$$
This map vanishes on $\N$: for $\l\in\N$ we have
$$
\norms{U_0\l}=\norms{\sum_{x\in\calS}\l(x)V(x)}
          =\sum_{x\in\calS}\sum_{y\in\calS}\bar{\l(x)}K(x,y)\l(y)
                        =\inp{\l}{\l}_\calL=0\;.
$$
So $U_0$ may be considered as a map $\calL/\N\to\calH$.
By the same calculation we find that $U_0$ is isometric.
Since $\bigvee V(\calS)$ is dense in $\calH$ and
$\bigvee V_K(\calS)$ is dense in $\calH_K$,
$U_0$ extends to a unitary map $U:\calH_K\to\calH$ mapping $V_K(x)$ to $V(x)$.
\end{proof}

\begin{examples}   \label{ex 1.7}
$\,$
\begin{alist}
\item
Let $\calS$ be any set and let $K(x,y):=\d_{x,y}$.
Then $\calH=l^2(\calS)$ and $V$ maps the elements of $\calS$ to
the standard orthonormal basis
of $\calH$.
\item
Let $\calS:=\calH_1\times\calH_2$,
the Cartesian product of two Hilbert spaces $\calH_1$ and $\calH_2$.
Let
$$
K\bigl((\psi_1,\psi_2),(\chi_1,\chi_2)\bigr)
        :=\inp{\psi_1}{\chi_1}\cdot\inp{\psi_2}{\chi_2}\;.
$$
Then $\calH=\calH_1\ten\calH_2$,
the tensor product of $\calH_1$ and $\calH_2$,
and $V(\psi_1,\psi_2)=\psi_1\ten\psi_2$.

\item
Let $\calS$ be a Hilbert space; call it $\K$ for the occasion.
Let $K(\psi,\chi):=\inp\psi\chi^2$.
Then $\calH$ is the {\it symmetric} tensor product $\K\sten\K$
and $V(\psi)=\psi\ten\psi$.

\item
Let $\calS:=\K$ as in Example (c).
Let $K$ be the positive definite kernel
$$
K(\psi,\chi):=e^{\inp\psi\chi}\;.
$$
Then the Kolmogorov dilation is the {\it Fock space} $\F(\K)$
over $\K$, defined as
$$
\F(\K):=\CC\oplus\K\oplus\half\bigl(\K\sten\K\bigr)\oplus
                         \sixth\bigl(\K\sten\K\sten\K\bigr)\oplus\cdots\cdots
$$
and $V(\psi)$ is the so-called {\it exponential vector} or
{\it coherent vector}
$$
\Exp(\psi):=1\oplus\psi\oplus(\psi\ten\psi)\oplus(\psi\ten\psi\ten\psi)\oplus
            \cdots
$$

\item
Let $\calS=\RR$ and let $K:\RR\times\RR\to\CC$ be given by
$$
K(s,t):=e^{-\eta|s-t|+i\o(s-t)}\;,\qquad(\eta>0,\o\in\RR).
$$
The Kolmogorov dilation of this kernel can be cast in the form
$$
\calH=L^2(\RR,2\eta dx)\;;\qquad
V:t\mapsto v_t\in L^2(\RR):\quad v_t(x)
                       :=\begin{cases}e^{(\eta-i\o)(x-t)}& \text{if } x\leq t;\\
                                                0& \text{if } x>t. \end{cases}
$$

\end{alist}
\end{examples}


\section{Some quantum mechanics}  \label{two}
Quantum mechanics is a physical theory that fits in the framework
of noncommutative probability, but which has much more structure.
It deals with particles and fields, using observables like position,
momentum, angular momentum, energy, charge, spin, isospin, etc.
All these observables develop in time according to a certain dynamical rule,
namely the Schr\"odinger equation.

In this section we shall pick out a few elements of this theory
that are of particular
interest to our main example: the damped harmonic oscillator
considered as a quantum Markov chain.

\subsection{Position and momentum}
Let us start with a simple example: a particle on a line.
This particle must have a {\it position} observable,
a projection valued measure on the Borel $\s$-algebra $\Si(\RR)$ of the
real line $\RR$:
   $$E:\Si(\RR)\to\B(\calH)\;.$$
The easiest choice (valid when the particle is alone in the world and has
no further degrees of freedom) is
   $$\calH:=L^2(\RR)\;;$$
   $$E(S):\psi\mapsto1_S\cdot\psi\;.$$
In this example the Hilbert space $\calH$ naturally carries a second real-valued
random variable in the form of the group
$(T_t)_{t\in\RR}$ of spatial translations:
\begin{equation} \label{foura}
   (T_t\psi)(x):=\psi(x-\hbar t)\;,
   \end{equation}
according to the remark following Stone's theorem (Theorem \ref{stone}).
This second observable is called the {\it momentum}
of the particle.
The constant $\hbar$ is determined by the units of length and of momentum
which we choose to apply.
The associated self-adjoint operators are
$Q$ and $P$ given by
\begin{align} \label{fivea}
(Q\psi)(x)&= x \psi(x)\;; \nonumber \\
(P\psi)(x)&=-i\hbar{\del\over{\del x}}\psi(x)\;.
\end{align}

Just as we have $T_t=e^{-itP}$,
it is natural to introduce $S_s:=e^{isQ}$ whose action on $\calH$ is
\begin{equation} \label{sixa}
S_s\psi(x):=e^{isx}\psi(x)\;.
\end{equation}
The operators $P$ and $Q$ satisfy Heisenberg's
{\it canonical commutation relation} (CCR)
\begin{equation} \label{seven}
[P,Q]=-i\hbar\cdot\one\;.
\end{equation}
A pair of self-adjoint operators $(P,Q)$ satisfying \eqref{seven} is called a
{\it canonical pair}.

\subsubsection{Representations of the canonical commutation relations}

What kinds of canonical pairs are there?

Before this question can be answered, it has to be reformulated.
Relation \eqref{seven}
 is not satisfactory as a definition of a canonical pair
since the domains on the left and on the right are not the same.
Worse than that,
quite pathological examples can be constructed,
even if \eqref{seven} is postulated to hold on a dense stable domain,
with the property that $P$ and $Q$ only admit unique
self-adjoint extensions (\cite{[ReS]}).

In order to circumvent such domain complications,
Weyl proposed to replace \eqref{seven} by a relation between
the associated unitary groups $(T_t)$ and $(S_s)$, namely:
\begin{equation} \label{eight}
T_tS_s=e^{-i\hbar st}S_sT_t\;,\qquad(s,t\in\RR)\;.
\end{equation}
It was von Neumann's idea to combine the two
into a two-parameter family
\begin{equation} \label{nine}
   W(t,s):=e^{{{i\hbar}\over2}st}T_tS_s\;,
   \end{equation}
forming a `twisted' representation of $\RR^2$,
as expressed by the {\it Weyl relation}: for all $s,t,u,v\in\RR$,
\begin{equation} \label{ten}
   W(t,s)W(u,v)=e^{-{{i\hbar}\over2}(tv-su)}W(t+u,s+v)\;.
   \end{equation}
This relation captures the group property of $T_t$ and $S_s$ together with the
relation \eqref{eight}.
Formally,
$$
W(t,s)=e^{i(sQ-tP)}\;.
$$
We shall call the representation on $L^2(\RR)$ of the CCR
given by \eqref{foura}, \eqref{fivea}, \eqref{sixa} and \eqref{nine} the
{\it standard representation} of the CCR.

Here and in the rest of the text we shall follow the quantum probabilist's
convention (\cite{[Mey]}), namely that
$$
\hbar=2\;.
$$


\begin{thm}[von Neumann's Uniqueness Theorem]
Let $\bigl(W(t,s)\bigr)_{t,s\in\RR}$ be a strongly
continuous family of unitary
operators on some Hilbert space $\calH$ satisfying the Weyl relation
\tu{\eqref{ten}}.
Then $\calH$ is unitarily equivalent with $L^2(\RR)\ten\K$,
such that $W(t,s)$ corresponds to $W_S(t,s)\ten\one$,
where $W_S$ is the standard representation 
of the CCR.
\end{thm}

\begin{proof}
Let $W:\RR^2\to\U(\calH)$ satisfy the Weyl relation \eqref{ten}.
For each integrable function
 $f:\RR^2\to\CC$ with $\int\int|f(t,s)|dt\,ds<\infty$,
define a bounded operator $A(f)$ on $\calH$ by
the strong sense integral
$$
A(f):=\intR\intR f(t,s)W(t,s)dt\,ds\;.
$$
We find the following calculating rules for such operators $A(f)$ and
their {\it kernels} $f$:
\begin{eqnarray*}
A(f)+A(g)&=&A(f+g)\;;\\
         A(f)^*&=&A(\widetilde f),\text{ where }
         \widetilde f(t,s):=\bar{f(-t,-s)}\;;\\
         A(f)A(g)&=&A(f*g)\;.
\end{eqnarray*}

Here the `twisted convolution product' $*$ is defined by
$$
(f*g)(t,s):=\intR\intR e^{-i(tv-su)}f(t-u,s-v)g(u,v)du\,dv\;.
$$
Moreover we claim that an operator (on a nontrivial Hilbert space) can
have at most one kernel:
\begin{equation} \label{eleven}
A(f)=0\Implies \calH=\{0\}\hbox{ or }f=0\;.
\end{equation}
Indeed,
if $A(f)=0$ then we have for all $a,b\in\RR$,
$$
0=W(a,b)^*A(f)W(a,b)=\intR\intR e^{2i(as-bt)}f(t,s)W(t,s)dt\,ds\;.
$$
Applying the linear functional $A\mapsto\inp\ph {A\psi}$ with $\ph ,\psi  \in\calH$
to both sides of this equation,
we find that for all $\ph ,\psi \in\calH$ the (integrable) function
$$
(t,s)\mapsto f(t,s)\inp{\ph}{W(t,s)\psi}
$$
has Fourier transform 0.
By the separability of $\calH$,
 either $W(t,s)=0$ for some $(t,s)$, (that is $\calH=\{0\}$),
or $f(t,s)=0$ for almost all $(t,s)$.

The key to the proof of uniqueness is the operator
$$
E:={1\over\pi}\intR\intR \gaus ts W(t,s)dt\,ds\;.
$$
It has the remarkable property that for all $a,b\in\RR$,
$EW(t,s)E$ is a scalar multiple of $E$:
\begin{equation} \label{twelve}
EW(a,b)E=\gaus ab E\;.
\end{equation}
Indeed, $E$ has kernel $g(t,s):={1\over\pi}\gaus ts$,
and the product $W(a,b)E$ has kernel
$$
h(u,v):={1\over\pi}e^{-i(av-bu)}\cdot\gaus{(a-u)}{(b-v)}\;.
$$
So $EW(a,b)E$ has kernel
\begin{eqnarray*}
(g*h)(t,s)&=&\int\int e^{-i(tv-su)}g(t-u,s-v)h(u,v)du\,dv\\
                   &=&{1\over{\pi^2}}\int\int e^{-i(tv-su)}
            \gaus{(t-u)}{(s-v)}e^{-i(av-bu)}\gaus{(a-u)}{(b-v)}du\,dv\\
                   &=&{1\over{\pi^2}}\gaus ab\gaus ts
            \int\int e^{-(u^2+v^2)}e^{(u-iv)(t+is+a+ib)}du\,dv\\
                   &=&{1\over{\pi^2}}\gaus ab\gaus ts
            \int\int e^{-\left(u-{1\over2}(t+is+a+ib)\right)^2
                        -\left(v-{i\over2}(t+is+a+ib)\right)^2}du\,dv\\
                   &=&{1\over\pi}\gaus ab\cdot\gaus ts\\
                   &=&\gaus ab g(t,s)\;,
\end{eqnarray*}
which proves \eqref{twelve}.

We conclude that $E^*=E$ (since $\widetilde g=g$),
$E^2=E$ (putting $a=b=0$ in \eqref{twelve}),
and that $E\A E=\CC E$,
where $\A$ is the von Neumann algebra generated by the Weyl operators.
So $E$ is a minimal projection in $\A$.
Denote its range by $\K$.
Then we have for all $\ph,\psi \in\K$ and all $t,s,u,v\in\RR$:
\begin{eqnarray*}
\inp{W(t,s)\ph}{W(u,v)\psi}&=&\inp{W(t,s)E\ph}{W(u,v)E\psi}\cr
           &=&\inp\ph{EW(-t,-s)W(u,v)E\psi}\cr
           &=&e^{i(tv-su)}\inp\ph{EW(u-t,v-s)E\psi}\cr
           &=&e^{i(tv-su)}\gaus{(u-t)}{(v-s)}\inp\ph{E\psi}\cr
           &=&e^{(t-is)(u+iv)}e^{-{1\over2}(t^2+s^2+u^2+v^2)}\inp\ph\psi\;,
\end{eqnarray*}
Therefore the map
$$
V:\RR^2\times\K\to\calH:
\bigl((t,s),\ph\bigr)\mapsto e^{{1\over2}(t^2+s^2)}W(t,s)\ph
$$
is a Kolmogorov dilation  (cf. Section \ref{one}) of the positive definite kernel
\begin{equation} \label{thirteen}
K:(\RR^2\times\K)\times(\RR^2\times\K)\to\CC ,   \quad
 \bigl((t,s),\ph;(u,v),\psi\bigr)\mapsto e^{(t-is)(u+iv)}\inp\ph\psi\;.
 \end{equation}
By explicit calculation you will find that $E_S$ is the orthogonal projection
onto the one-dimensional subspace spanned by the unit vector
$\O(x):=\sqrt{\g(x)}$, where
$$
\g(x):={1\over{\sqrt{2\pi}}}e^{-{1\over2}x^2}\;.
$$
So in the standard case the dilation is
$$
V_S:\RR^2\to L^2(\RR):(t,s)\mapsto e^{{1\over2}(t^2+s^2)}
                                   e^{-{i\over2}ts}e^{isx}\O(x-2t)\;.
$$
By Kolmogorov's Dilation Theorem,
there exists a unitary equivalence $U:L^2(\RR)\ten\K\to\calH$
such that for all $a,b\in\RR$ and $\psi\in\K$:
$$
U\bigl(W_S(a,b)\O\ten\psi\bigr)=W(a,b)\psi\;,
$$
and therefore for all $a,b\in\RR$:
$$
W(a,b)=U\bigl(W_S(a,b)\ten\one\bigr)U^{-1}\;,
$$
provided that the range of $V$ is dense in $\calH$.
Let $\calL$ denote the orthogonal complement of this range.
Then $\calL$ is invariant for the Weyl operators;
let $W_0(t,s)$ be the restriction of $W(t,s)$ to $\calL$.
Construct $E_0:=A_0(g)$ in terms of $W_0$ in the same way as $E$ was
constructed from $W$.
Then clearly $E_0\le E$, but also $E_0\perp E$.
So $E_0=A_0(g)=0$ and by \eqref{eleven} we have $\calL=\{0\}$.
\end{proof}

\begin{exercise}
Calculate the minimal projection $E_S$ in the standard representation.
\end{exercise}

\subsection{Energy and time evolution}
The evolution in time of a closed quantum system is given by a
pointwise strongly continuous one-parameter group $(\a_t)_{t\in\RR}$ of
*-auto\-morphisms of the observable algebra $\A$.

Like in the case of a particle on a line,
for a finite number $n$ of (distinguishable) particles in
$d$-dimensional space we take $\A=\B(\calH)$ with $\calH=L^2(\RR^{nd})$.
Since all automorphisms of this algebra are implemented by unitary
transformations of $\calH$, the group $(\a_t)$ is of the form
   $$\a_t(A)=U_tAU_t^{-1}\;.$$
It is possible to choose the unitaries so that
$t\mapsto U_t$ is a strongly continuous unitary representation $\RR\to\U(\calH)$.
We denote its Stone generator by $H/\hbar$:
   $$U_t=e^{itH/\hbar}\;.$$
The self-adjoint operator $H$ corresponds to an observable of the system
of particles, called its {\it energy}.
The operator $H$ itself is known as the {\it Hamilton operator}
or {\it Hamiltonian} of the system.
As the Hamiltonian commutes with the time evolution operators,
energy is a conserved quantity:
   $$\a_t(H)=U_tHU_t^{-1}=H\;.$$
The nature of a physical system is characterised by its
{\it dynamical law} (a term of Hermann Weyl, see \cite{[Wey]}).
This is an equation which expresses the Hamiltonian in terms of other
observables.
For $n$ interacting particles in $\RR^d$ in
the absence of magnetic fields the dynamical law takes the form
   $$H=\sum_{j=1}^{nd}{1\over{2m_{k(j)}}}P_j^2+V(\tuple Q_{nd})\;$$
for some function $V:(\RR^d)^n\to\RR$, called the {\it potential}.
The positive constants $m_k$, $k=1,\cdots,n$ are the {\it masses} of the
particles.
(Incidentally we put $k(j):=1+[(j-1)/d]$,
where $[\,\,\,]$ denotes integer part,
 in order to attach the same
mass to the coordinates of the same particle.)

\subsubsection{Free particles}
If $V=0$,
then $U_t$ factorises into a tensor product of $nd$ one-dimensional
evolution operators,
all of the form
   $$U_t=e^{itH/\hbar}=e^{i{t\over{2m\hbar}}P^2}\;.$$
Since the Hamiltonian $H=P^2/2m$ now commutes with $P$,
momentum is conserved:
   $$\a_t(P)=P\;.$$
On a formal level the time development of the operator $Q$ is
found by solving the differential equation
\begin{equation} \label{fourteen}
 \dd t\a_t(Q)=\dd tU_tQU_t^{-1}={i\over{2m\hbar}}[P^2,\a_t(Q)],
 \end{equation}
a solution of which is
   $$\a_t(Q)=Q+{t\over m}P\;.$$
According to the Uniqueness Theorem the canonical pairs $(P,Q)$ and
$(P,Q+{t\over m}P)$ are indeed unitarily equivalent.
So we expect the evolution of the Weyl operators to be the
following:
\begin{align*}
 \a_t\bigl(W(x,y)\bigr)&=\a_t\left(e^{-ixP+iyQ}\right)
                           =e^{-ixP+iy(Q+{t\over m}P)}\\
                          &=e^{-i(x-{t\over m}y)P+iyQ}
                           =W\left(x-{t\over m}y\;,y\right)\;.
\end{align*}

\begin{propn}\label{P 2.2}
Let $P:=-i\hbar{\del\over{\del x}}$ denote the momentum operator on
$\calH:=L^2(\RR)$,
and let $W:\RR^2\to\U(\calH)$ be given by \eqref{ten}.
Let
   $$U_t:=e^{i{t\over{2m\hbar}}P^2}\;.$$
Then
   $$U_tW(x,y)U_t^{-1}=W\left(x-{t\over m}y\;,y\right)\;.$$
\end{propn}

\begin{proof}
From the definitions of $T_t$ and $E_Q$ it follows that
for all measurable sets $B\subset\RR$ and all $t\in\RR$:
$$
T_tE_Q(B)T_t^{-1}=E_Q(B+\hbar t)\;.
$$
By the uniqueness theorem irreducible representations
of the CCR have the symmetry $Q\to P$, $P\to-Q$.
So we also have the exchanged imprimitivity relation
   $$\forall_{B\in\Si(\RR)}\forall_{y\in\RR}:\quad
             S_yE_P(B)S_y^{-1}=E_P(B+\hbar y)\;.$$
Hence for all $y,t\in\RR$,
\begin{align*}
   S_yU_{-t}S_y^{-1}
    &=S_y\left(\intR e^{-i{t\over{2m\hbar}}\l^2}E_P(d\l)\right)S_y^{-1}\cr
    &=\intR e^{-i{t\over{2m\hbar}}(\l-\hbar y)^2}E_P(d\l)\cr
    &=U_{-t}\cdot T_{-{t\over m}y}\cdot e^{-i{{t\hbar}\over{2m}}y^2}\;.
\end{align*}
Multiplying by $U_t$ on the left and by $S_y$ on the right we find
   $$U_tW(0,y)U_t^{-1}=U_tS_yU_t^{-1}
                      =e^{-i{{t\hbar}\over{2m}}y^2}T_{-{t\over m}y}S_y
                      =W\left(-{t\over m}y\;,y\right)\;.$$
As $T_x$ commutes with $U_t$ we may freely add $(x,0)$ to the argument
of $W$, and the proposition is proved.
\end{proof}

By imposing some state $\ph$ on $\A=\B(L^2(\RR))$,
all stochastic information on the model $(\A,\ph,\a_t)$ can be obtained
from the evolution equation $\a_t(Q)=Q+{t\over m}P$.
For example, at large times $t$ the random variable
${1\over t}\a_t(Q)$ approaches
${1\over m}P$ in distribution, provided that $\ph$ does not favour
large $Q$ values too much.
So a position measurement at a late time can serve as a measurement
of momentum at time 0.
This puts into perspective the well-known uncertainty principle
for position and momentum at equal times.

\subsubsection{The Schr\"odinger picture and the Schr\"odinger equation}
The type of description of a system given so far,
namely with random variables moving in time,
and the state $\ph$ given once and for all,
is called the {\it Heisenberg picture} of quantum mechanics.
In probability theory this is common usage,
and we shall adopt it also in quantum probability.

However, quantum mechanics is often thought of in a different way,
where one lets the state move, and keeps the operators fixed.
This is close to Schr\"odinger's `wave mechanics',
and is therefore called the {\it Schr\"odinger picture}:

If we take for $\ph$ a pure (that is extremal) state on the algebra
$\A=\B(\calH)$ where $\calH$ is, say, $L^2(\RR^{nd})$:
    $$\ph(A)=\inp{\psi}{A\psi}\;,\qquad(\psi\in\calH,\norm\psi=1),$$
then we can express all probabilities at later times $t$ in terms of
the {\it wave function}
    $$\psi(\tup x_{nd};t):=(U_t^{-1}\psi)(\tup x_{nd})\;.$$
This wave function satisfies the {\it Schr\"odinger equation},
a partial differential equation reflecting the dynamical law:
\begin{align*}
&-i\hbar{\del\over{\del t}}\psi(\tup x_{nd};t)\\
   &=\sum_{j=1}^{nd}-{1\over{2m_{k(j)}\hbar^2}}{{\del^2}\over{\del x_j^2}}
        \psi(\tup x_{nd};t)+V(\tup x_{nd})\psi(\tup x_{nd};t)\;.
\end{align*}
If $E$ is an orthogonal projection in $\calH$,
then the probability of the associated event can be calculated in the
Schr\"odinger picture by
\begin{align*}
 \ph(\a_t(E))&=\inp{\psi}{U_tEU_t^{-1}\psi}
                           =\inp{U_t^{-1}\psi}{EU_t^{-1}\psi}\\
    &=\int_{\RR^{nd}}\overline{\psi_t(\tup x_{nd})}
                     \bigl(E\psi_t\bigr)(\tup x_{nd})\tup{dx}_{nd}\;.
\end{align*}

\subsection{The harmonic oscillator}
A harmonic oscillator is a canonical pair $(Q,P)$ of observables that under
time evolution $(\a_t)_{t\in\RR}$ performs a rotation such as
\begin{align*}
  \a_t(Q)&=Q\cos t+P\sin t\;;\\
              \a_t(P)&=-Q\sin t+P\cos t\;.
\end{align*}
Since rotation in the plane is symplectic (preserves the area two-form),
this evolution respects the canonical commutation relation
$QP-PQ=i\hbar\cdot\one$.
So by the Uniqueness Theorem it determines
(up to a time-dependent phase) a group of unitary transformations
$(U_t)_{t\in\RR}$ of the Hilbert space on which it is represented.
(For example,
$U_{\pi\over2}$ is a unitary transformation of $L^2(\RR)$
that sends $Q$ into $P$ and $P$ into $-Q$ in the standard representation
\eqref{fivea}. This is the Fourier transform.)

Making a formal calculation as in \eqref{fourteen}
by differentiating the equality
   $$\a_t(A)=e^{itH/\hbar}Ae^{-itH/\hbar}$$
we find that a Hamiltonian of the form
\begin{equation} \label{Hamharmosc}
   H=\half(P^2+Q^2)
   \end{equation}
can be expected to generate such a rotating evolution.

\smallskip\noindent
The textbook treatment of the harmonic oscillator (e.g. \cite{[Han]}),
follows the elegant algebraic reasoning of Dirac,
who rewrote the Hamiltonian \eqref{Hamharmosc} as
   $$H=\half(Q-iP)(Q+iP)+\half i[P,Q]=:\hbar a^*a+\half\hbar\cdot\one\;.$$
The operators $a$ and $a^*$ are then seen to lower and raise the
eigenvalue of $H$,
and are called the {\it annihilation} and {\it creation} operators.

Here we choose to proceed more analytically,
seizing the opportunity to introduce techniques which will
be useful again later on for the treatment of free quantum fields
and the damped oscillator.

Our goal is to describe $H$ and $U_t$ explicitly.

\subsubsection{Heisenberg's matrix representation}

First we note that,
since $\a_t$ has period $2\pi$,
the differences between spectral points of $H$ must be multiples
of $\hbar$.
On the grounds of \eqref{Hamharmosc} we suspect that $H$ is bounded from below,
so let us try
   $$\sp(H)=\hbar\NN+c\;.$$
We take as our Hilbert space
$\calH_H:=l^2\left(\NN,{\textstyle{1\over{n!}}}\right)$
with the Hamiltonian given by
   $$(H\th)(n)=(\hbar n+c)\th(n)\;.$$
The subscript `$H$' indicates that on this space we wish to stage
matrix mechanics of the Heisenberg type.
If we define on $\calH_H$ the `product' or `coherent' vectors
   $$\pi(z):=(1,z,z^2,z^3,\cdots),\quad(z\in\CC)\;,$$
then our intended time evolution takes the form
\begin{equation} \label{UHt}
   U^H_t\pi(z)=e^{itc/\hbar}\pi\left(e^{it}z\right)\;.
   \end{equation}
Now we want to represent a canonical pair $(P,Q)$ in this space,
or equivalently, Weyl operators $W(z)$, that rotate in the same way:
$U_tW(z)U_t^{-1}=W(e^{it}z)$.
We note that
   $$\inp{\pi(u)}{\pi(v)}=\szi n{{\bar u^nv^n}\over{n!}}=e^{\bar uv}\;,$$
so that we have here another dilation of the positive definite kernel
\eqref{thirteen} used in the proof of the Uniqueness Theorem.
An irredicible representation of the CCR is close at hand. Put:
   $$W_H(z)\pi(u)=e^{-\bar zu-{1\over2}|z|^2}\pi(u+z)\;,\quad(z,u\in\CC)\;.$$
These operators satisfy the Weyl relation
\begin{equation} \label{WeylC}
   W_H(w)W_H(z)=e^{-i\Im(\bar wz)}W_H(w+z)\;,
   \end{equation}
the same as \eqref{ten} if we identify $W(t,s)$ with $W_H(t+is)$.
Clearly we have also obtained
\begin{equation} \label{WeylCtime}
   U_tW(z)U_t^{-1}=W\left(e^{it}z\right)\;.
   \end{equation}
Let us summarise, again replacing $\hbar$ by 2.

\begin{propn} \label{P 2.3}
The Heisenberg representation of the Harmonic oscillator is given by
\begin{align*}
 \calH_H&=l^2(\NN,{1\over{n!}})\;;\cr
 (H_H\th)(n)&=(2n+1)\th(n);\qquad U_t^H\pi(z)=e^{{i\over2}t}\pi(e^{it}z)\;;\\
 W_H(z)\pi(u)&=e^{-\bar zu-{1\over2}|z|^2}\pi(u+z)\;.
 \end{align*}
In concrete terms, on the standard orthonormal basis,
   $$Q=\begin{bmatrix}0&1&0&0&0&\ldots\\
                     1&0&\sqrt2&0&0&\ldots\\
                     0&\sqrt2&0&\sqrt3&0&\ldots\\
                     0&0&\sqrt3&0&\sqrt4&\ldots\\
                     0&0&0&\sqrt4&0&\ldots\\
                     \vdots&\vdots&\vdots&\vdots&\vdots&\ddots \end{bmatrix},
\quad P={1\over i}\begin{bmatrix}0&1&0&0&0&\ldots\\
                     -1&0&\sqrt2&0&0&\ldots\\
                     0&-\sqrt2&0&\sqrt3&0&\ldots\\
                     0&0&-\sqrt3&0&\sqrt4&\ldots\\
                     0&0&0&-\sqrt4&0&\ldots\\
                     \vdots&\vdots&\vdots&\vdots&\vdots&\ddots \end{bmatrix}\;.$$
These matrices satisfy
   $$QP-PQ=2i\cdot\one  \tu{\text{ and }} \, \half(Q^2+P^2)=H\;,$$
where
   $$H=\begin{bmatrix}1&&&&&\\
                     &3&&&\emptyset&\\
                     &&5&&&\\
                     &&&7&&\\
                     &\emptyset&&&9&\\
                     &&&&&\ddots \end{bmatrix}\;.$$
\end{propn}

\begin{proof}
It only remains to check the matrices for $Q$ and $P$.
We note that
   $$e^{iyQ}\pi(u)=W_H(iy)\pi(u)=e^{iyu-{1\over2}y^2}\pi(u+iy)\;,$$
and we find by differentiation
   $$Q\pi(u)=u\pi(u)+\pi'(u)\;.$$
Taking the coefficient of $u^n$ the matrix of $Q$ is found.
The matrix for $P$ is found in the same way.
The choice of the ground state energy $c=\half\hbar=1$ in the definition of
$H$ fixes the relation with $Q$ and $P$ correctly.
\end{proof}

\subsubsection{The Gaussian representation}
Here is another useful representation of the harmonic oscillator algebra
on a Hilbert space.

Let $\calH_G:=L^2(\RR,\g)$, where $\g$ is the standard Gauss measure on $\RR$:
   $$\g(dx):=\g(x)dx:={1\over\sqrt{2\pi}}e^{-\hf x^2}\,dx\;.$$
Define for $z\in\CC$ the vector $\e(z)$ by
   $$\e(z):x\mapsto e^{zx-\hf z^2}\;.$$
Then $\e(z)$ with $z\in\CC$ is a total set in $\calH_G$.
(Actually, $z\in i\RR$ is already sufficient by the uniqueness of the Fourier
transform.)
Again we find
   $$\inp{\e(z)}{\e(u)}=e^{\bar zu}\;,\qquad(z,u\in\CC)\;.$$

\begin{propn} \label{P 2.4}
There exists a unitary map $U_{HG}:\calH_H\to\calH_G$ such that for all $z\in\CC$
   $$U_{HG}\,\pi(z)=\e(z)\;.$$
This map sends the vector $e_n:=(0,\cdots,0,1,0,\cdots)$ into the $n$-th
Hermite polynomial, where these polynomials are given by the generating
function
   $$\szi n z^nh_n(x)=e^{zx-\hf z^2}\;.$$
Consequently, this version of the Hermite polynomials satisfies
   $$\intR h_n(x)h_m(x)\,\g(dx)={1\over{n!}}\d_{nm}\;.$$
\end{propn}

\begin{proof}
The map $\pi(z)\mapsto\e(z)$ extends to a unitary map since
the linear spans of the ranges of $\pi$ and $\e$
are dense and both $\pi$ and $\e$ are minimal
dilations of the positive definite
kernel $(z,u)\mapsto e^{\bar zu}$.
\end{proof}

Let us carry over the relevant operators with this unitary transformation.
We find:
   $$\left(e^{isQ_G}\psi\right)(x)=e^{isx}\psi(x),
                  \quad (Q_G\psi)(x)=x\psi(x)\;;$$
   $$\left(e^{-itP_G}\psi\right)(x)
                  =\psi(x-2t)\left({{\g(x-2t)}\over{\g(x)}}\right)^{1/2},
                  \quad (P_G\psi)(x)=ix\psi(x)-2\psi'(x)\;;$$
   $$H_G\psi=(2N_G+\one)\psi
      =-2{{\del^2}\over{\del x^2}}\psi+2x{\del\over{\del x}}\psi+\psi\;.$$

\subsubsection{The Schr\"odinger representation}
Finally we get to the standard Schr\"odinger representation
\eqref{foura}, \eqref{sixa} and \eqref{nine} of the harmonic
oscillator by dividing away a factor $\sqrt{\g(x)}$.
Let $\calH_S:=L^2(\RR)$ and define
   $$U_{GS}:\calH_G\to\calH_S:(U_{GS}\psi)(x):=\sqrt{\g(x)}\psi(x)\;.$$

\subsection{The problem of damping} \label{sub 2.4}
A damped harmonic oscillator is an evolution $(T_t)_{t\ge0}$
on the real-linear span of a canonical pair $(P,Q)$ that has the form
\begin{align} \label{damposc}
 T_t(Q)&=e^{-\eta t}\left(Q\cos\o t+P\sin\o t\right)\;,\nonumber \\
             T_t(P)&=e^{-\eta t}\left(-Q\sin\o t+P\cos\o t\right)\;,
                    \qquad(\eta>0)\;.
\end{align}

(We apologise for a clash of notation: $T_t$ is not related to
translations.)
This spiralling motion in the plane compresses areas by a factor $e^{-2\eta
t}$,
so that for $t>0$ the operators $T_t(Q)$ and $T_t(P)$ disobey the
canonical commutation relation,
and $T_t$ cannot be extended to an automorphism of $\B(\calH)$.

Yet this damped oscillatory behaviour occurs in nature,
for instance when
an atom is loosing its energy to its surroundings by emission of
light.
So it would be worth while to make sense of it.
There are two basic questions related to this model.

\smallskip
\noindent \textbf{Question 1}.
How should $T_t$ be extended to $\B(\calH)$?

\noindent \textbf{Question 2}.
Can $(T_t)_{t\ge0}$ be explained as part of a larger whole that evolves
by  \newline
*-automorphisms of the form $a_t (A) = U_t AU^{-1}_t$,
where $U^{-1}_t$ satisfies a Schr\"odinger equation?

\smallskip

\subsubsection{Spirals and jumps}
In Heisenberg's matrix mechanics atoms were supposed to move in a mixture
of two ways.
Most of the time they were thought to rotate according to the evolution
$U_t^H$ as described above,
but occasionally they made random jumps down the ladder of eigenvalues
of the energy operator $H$.
Each time an atom made such a jump,
it emitted a quantum of light whose (angular) frequency $\o$ was related
to the size $E$ of the jump by
   $$E=\hbar\o\;.$$
The probability per unit of time for the atom to jump was given by Fermi's
`Golden Rule', formulated in terms of the coupling between the atom and
its surroundings, and it is proportional to the damping rate $\eta$.

In the following sections we shall describe this behaviour as a
quantum Markov process.
Both jumps and spirals will be visible in the extension of our $T_t$
to the atom's full observable algebra.
This will be our answer to Question 1,
for which we shall need the notion of completely positive operators.

Our answer to Question 2 will be a reconstruction of the atom's surroundings:
a {\it dilation}.
There we shall see how the atom can absorb and emit quanta.

\section{Conditional expectations and operations} \label{three}

We shall now give a sketch of the operational approach to quantum
probability which was pioneered by Davies, Lewis and Evans
(\cite{[Dav]}, \cite{[EvL]}).

\subsection{Conditional expectations in finite dimension}

In this section we choose for definiteness: $\A:=M_n$,
the algebra of all complex $n\times n$ matrices, and
   $$\ph:\A\to\CC:\quad A\mapsto\tr(\r A)\;,$$
where $\r$ is a symmetric $n\times n$ matrix with strictly positive
eigenvalues and trace 1, so that $\ph$ is faithful.

Let $A$ be a symmetric $n\times n$ matrix with the spectral decomposition
   $$A=\ssp\a A:\a E_\a\;.$$
The orthogonal projections $E_\a$, $\a\in\sp(A)$,
form a partition of unity.
{\it Measuring} the {\it observable} $A$ means asking all the compatible
questions $E_\a$ at the same time.
Precisely one of the answers will be `yes',
as stipulated in the interpretation rules.
If the answer to $E_\a$ is `yes', then $A$ is said to take the value $\a$.
This happens with probability $\ph(E_\a)$.

It is natural to define the {\it expectation} of $A$ as
   $$\ssp\a A:\a\ph(E_\a)=\ph\left(\ssp\a A:\a E_\a\right)=\ph(A)\;.$$
So the state $\ph$ not only plays the role of a probability measure,
but naturally extends to the associated expectation.

\smallskip\noindent
Now let $B=B^*\in\A$ be a second observable with spectral decomposition
   $$B=\ssp\b B:\b F_\b\;.$$
If we first measure $B$ and then $A$ in each trial,
in the limit of increasingly many trials
 we obtain a probability measure $\PP$ on $\sp(A)\times\sp(B)$.
By  the discussion of interpretation  of quantum probability in Section~\ref{one}
the probabilities are given by
   $$\PP(\{(\a,\b)\})=\ph(F_\b E_\a F_\b)\;.$$
It is then natural to define the conditional probability
$\PP[A=\a|B=\b]$ as that proportion of the trials that have yielded $B=\b$
which turn out to give $A=\a$ later:
   $$\PP[A=\a|B=\b]:={{\PP(\{(\a,\b)\})}\over{\ssp\a A:\PP(\{(\a,\b)\})}}
                    ={{\ph(F_\b E_\a F_\b)}\over{\ph(F_\b)}}\;.$$
The associated conditional expectation is naturally defined as
   $$\EE(A|[B=\b]):=\ssp\a A:\a\PP[A=\a|B=\b]
                   ={{\ph(F_\b A F_\b)}\over{\ph(F_\b)}}\;,$$
Note that this is a function, $f$ say, of $\b$.
Seen as a quantum random variable this conditional expectation is
described by the matrix $f(B)$:
\begin{equation} \label{CEone}
   \EE(A|B):=f(B)=\ssp\b B:f(\b)F_\b
              =\ssp\b B:{{\ph(F_\b A F_\b)}\over{\ph(F_\b)}}F_\b\;.
\end{equation}
Note that
   $$\ph(\EE(A|B))=\ssp\b B:\ph(F_\b AF_\b)\;.$$

\begin{rem}
In general we do {\it not} have
\begin{equation} \label{compphi}
   \ph(\EE(A|B))=\ph(A)\;.
\end{equation}
The left hand side is the expectation of $A$ after measuring $B$.
The right hand side is the expectation of $A$ without any previous operation.
The fact that these two expectation values can differ is typical
for quantum probability.
\end{rem}

Let us give a simple counterexample to \eqref{compphi} here:
Let $\A:=M_2$, choose $\l\in(0,1)$, and put
   $$\r=\begin{bmatrix}\l&0\\ 0&1-\l \end{bmatrix},\quad
             A=\begin{bmatrix}1&0\\ 0&0 \end{bmatrix},\quad
             B=\half\begin{bmatrix}1&1\\ 1&1 \end{bmatrix}\;.$$
It is readily checked that
   $$\ph(A)=\l,\qquad\ph\bigl(\EE(A|B)\bigr)=\half\;,$$
so that the equality \eqref{compphi}
holds if and only if $\ph$ is the trace state.

\subsubsection{The conditional expectation given a discrete random variable}

Let $\B$ denote the (abelian) subalgebra of $M_n$ generated by $\one$ and $B$,
a hermitian matrix.
In quantum probability theory it is accepted practice (\cite{[EvL]})
 to call $\EE(A|B)$
the {\it conditional expectation of $A$ given the algebra $\B$},
written $P_\B(A)$
{\it only} if the equality \eqref{compphi} {\it does} hold.
The reason is that the definition as it stands
does not generalise to observables $B$ with continuous spectrum,
or to noncommutative subalgebras $\B$ of $\A$.
(The value of $\ph(\EE(A|B))$ changes
if the possible values of $B$  are not all distinguished
while measuring $B$: if $\b_1$ and $\b_2$ are not distinguished,
their eigenspaces group together into a single subspace,
and in \eqref{CEone} the projections $F_{\b_1}$ and $F_{\b_2}$ are replaced by
the projection $F_{\b_1}+F_{\b_2}$.)

Note that the projections in $\B$ are labeled by {\it subsets}
of $\sp(B)$:
   $$\E(\B)=\set{\sum_{\b\in V}F_\b}{V\subset\sp(B)}\;,$$
and that $\B$ is the linear span of the projections $F_\b$.


\smallskip\noindent
The following is a finite dimensional version of Takesaki's theorem (\cite{[Tak]})
on the existence of conditional expectations onto von Neumann subalgebras.

\begin{thm} \label{T 3.1}
Let $B=B^*\in M_n$ and let $\B$ be the *-algebra generated by $\one$ and $B$.
Let $\ph :M_n\to\CC:A\mapsto\tr(\r A)$ with $\r$ strictly positive and
$\tr(\r)=1$.
Then the following are equivalent.
\begin{alist}
\item
There exists a linear map $P:M_n\to\B$ such that
\begin{equation} \label{CEtwo}
\forall_{A\in M_n}\forall_{F\in\E(\B)}:\quad\ph(FAF)=\ph(FP(A)F)\;.
   \end{equation}

\item There exists a linear map $P$ from $M_n$ onto $\B$ such that

\begin{rlist}

\item
$P$ maps positive definite matrices to positive definite matrices.
\item
$P(\one)=\one$;
\item
$\ph\circ P=\ph$;
\item
$P^2=P$.
\end{rlist}
\item
$B\r=\r B$.
\end{alist}

\end{thm}

If these equivalent conditions hold, then the linear maps $P$ mentioned
in (a) and (b) are the same.
It is called the {\it conditional expectation onto $\B$ compatible with}
$\ph$.

\begin{proof}
(a) $\implies$ (b):
suppose $P:M_n\to\B$ is such that \eqref{CEtwo} holds.

Let $A\ge0$ and decompose $P(A)$ as
$\ssp\b B:a_\b F_\b$ with $F_\b\in\E(\B)$.
Then $a_\b\ph(F_\b)=\ph(F_\b P(A))=\ph(F_\b P(A)F_\b)=\ph(F_\b A F_\b)\ge0$.
So $a_\b\ge0$ for all $\b$ and $P(A)\ge0$.

Putting $A=\one$ in \eqref{CEtwo} we find that for all $\b\in\sp(B)$:
$\ph(F_\b)=\ph(F_\b P(\one)F_\b)=\ph(F_\b P(\one))$.
Writing $P(\one)=\ssp\b B:e_\b F_\b$,
we see that $e_\b=1$,
hence $P(\one)=\one$.

By putting $F=\one$ in \eqref{CEtwo}, (iii) is obtained.

Finally, given $A$, the element $P(A)$ of $\B$ is obviously uniquely determined
by \eqref{CEtwo}. But if $A\in\B$, then $P(A):=A$ clearly satisfies
\eqref{CEtwo}. It follows that $P$ is an idempotent with range $\B$.

(b) $\implies$ (c):
Make a Hilbert space out of $\A=M_n$ by endowing it with the inner product
   $$\inpph{X}{Y}:=\ph(X^*Y)\;.$$
We claim that on this Hilbert space $P$ is an orthogonal projection.
Since $P$ is idempotent by assumption (b)(iv),
it suffices to show that $P$ is a contraction:
\begin{equation} \label{contraction}
\normph{P(A)}\le\normph{A}\;.
   \end{equation}
Given $A\in M_n$, define numbers $a_\b\in\CC$ and $b_\b\ge0$ by
   $$P(A)=\ssp\b B:a_\b F_\b\;;\qquad P(A^*A)=\ssp\b B:b_\b F_\b\;.$$
Then from the positivity property (b)(i) it follows that
   $$\forall_{\l\in\CC}:\quad
       P\bigl((\l\cdot\one-A)^*(\l\cdot\one-A)\bigr)\ge0\;.$$
This implies that for all $\b\in\sp(B)$ and all $\l\in\CC$,
   $$|\l|^2-(\bar\l a_\b+\l\bar{a_\b})+b_\b\ge0\;,$$
from which it follows that
   $$|a_\b|^2\le b_\b\;,\quad\hbox{that is}\quad P(A)^*P(A)\le P(A^*A)\;.$$
Applying $\ph$ to the last inequality and using (iii)
yields the statement \eqref{contraction}.
So $P$ is an orthogonal projection $M_n\to\B$, that is for all $A\in M_n$,
   $$A-P(A)\perp_\ph\B\;.$$
This means that for all $A\in M_n$:
   $$\ph(AB)=\ph(P(A)B)\, \text{ and } \, \ph(BA)=\ph(BP(A))\;.$$
But then, since $\B$ is commutative,
   $$\ph(BA)=\ph(BP(A))=\ph(P(A)B)=\ph(AB)\;.$$
It follows that
   $$\tr(\r BA)=\tr(\r AB)=\tr(B\r A)\;,$$
and (c) is proved.

(c) $\implies$ (a):
Suppose that $B\r=\r B$.
Then for all $F\in\E(\B)$ and all $A\in M_n$,
  $$\ph(FAF)=\tr(\r FAF)=\tr(F\r FA)=\tr(\r F^2A)=\tr(\r FA)=\ph(FA)\;.$$
Therefore, defining $P(A)$ by the r.h.s. of \eqref{CEone},
and putting $F=\sum_{\b\in V}F_\b$ with $V\subset\sp(B)$:
\begin{align*}
 \ph(FP(A)F)
          &=\ssp\b B:{{\ph(F_\b AF_\b)}\over{\ph(F_\b)}}\ph(F F_\b F)
               =\sum_{\b\in V}\ph(F_\b AF_\b)\\
          &=\sum_{\b\in V}\ph(F_\b A)=\ph(FA)
               =\ph(FAF)\;.
\end{align*}
\end{proof}


\subsection{Operations in finite dimension}

Let $\A$ and $\B$ be finite dimensional von Neumann algebras,
and let $\A^*$ and $\B^*$ denote their duals.
A linear map $T:\A\to\B$ defines by duality a linear map
$T^*:\B^*\to\A^*$.

The map $T^*$ maps states into states if and only if $T$ is {\it
positive},
that is maps positive elements of $\A$ to positive elements of $\B$,
and is identity preserving.

The map $T$ is said to be {\it n-positive} if
$T\ten\id$ maps positive elements of
$\A\ten M_n$ to positive elements of $\B\ten M_n$:
$$
\left(A_{ij}\right)_{i,j=1}^n\ge0\Implies
\left(T(A_{ij})\right)_{i,j=1}^n\ge0\;.
$$
$T$ is called {\it completely positive}  if it is $n$-positive for all
$n\in\NN$.
In that case $T^*\ten\id$ maps states on $\B\ten M_n$ to states on
$\A\ten M_n$.
$T$ is called {\it identity preserving} if $T(\one_\A)=\one_\B$.

\begin{defn}
An {\it operation} $T:\A\to\B$ is a completely positive identity
preserving map.
Adjoints of operations will also be called operations.
\smallskip

The idea is that any physical procedure which takes as an input a state on
some quantum system described by $\B$,
and which turns out a state on a quantum system described by $\A$
must necessarily be of the above kind.
Not all operations in the sense of the definition can actually be performed,
but certainly nothing else is physically possible.
Indeed any physical operation on a quantum system $\A$ should also
define a physical operation on $\A\ten\R$,
where $\R$ stands for some quantum system not affected by the operation.
The existence of such an `innocent bystander' outside our quantum system
$\A$ should never lead to the prediction by quantum theory of
negative probabilities.
\end{defn}

The following example shows that complete positivity is strictly stronger
than positivity.
Let
$$
T:M_2\to M_2:\begin{bmatrix}a&b\\ c&d \end{bmatrix}\mapsto
             \begin{bmatrix}a&c\\ b&d \end{bmatrix}\;.
$$
Then $T(A^*A)=T(A)T(A)^*\ge0$ for all $A$, but
$$
T\ten\id:M_2\ten M_2\to M_2\ten M_2 \text{ maps }
\begin{bmatrix}1&0&0&1\cr
                                                 0&0&0&0\cr
                                                 0&0&0&0\cr
                                                 1&0&0&1\end{bmatrix}
                                                 \text{ to }
                                \begin{bmatrix}1&0&0&0\cr
                                                 0&0&1&0\cr
                                                 0&1&0&0\cr
                                                 0&0&0&1\end{bmatrix} ;
$$
i.e. it maps a
 one-dimensional projection to a matrix with eigenvalues $1$ and $-1$.

\subsection{Operations on quantum probability spaces}

A quantum probability space $(\A,\ph)$ has a canonical representation
on a Hilbert space, called the GNS representation after Gel'fand,
Naimark and Segal.
It is the representation of $\A$ on $\calH_\ph$,
the Kolmogorov dilation of the positive definite kernel
$$
\A\times\A\to\CC:\quad(A,B)\mapsto\ph(A^*B)\;.
$$
States which are given by density matrices on this space are called
{\it normal states} on $\A$,
and the set of all normal states is denoted by $\A_*$.

When we write $T:(\A,\ph)\to(\B,\psi)$,
we mean that $T$ is a completely positive operator $\A\to\B$ such that
$T(\one_\A)=\one_\B$ and {\it also} $\psi\circ T=\ph$.
The latter condition, which can equivalently be written as
$$
T^*\psi=\ph\;,
$$
ensures that $T^*$ maps normal states to normal states.
This property is only relevant for infinite dimensional von Neumann
algebras.
When speaking of operations between quantum probability spaces we shall
always imply that the state is preserved.

\subsection{Quantum stochastic processes}

Let us now consider the category QP whose objects are
quantum probability spaces and whose morphisms are operations.

\begin{lem}[Schwartz's inequality for completely positive operators]  \label{L 3.2}
Let $T:(\A,\ph)\to(\B,\psi)$.
Then for all $A\in\A$,
   $$T(A^*A)\ge T(A)^*T(A)\;.$$
\end{lem}

\begin{proof}
Let $\A$ be represented on $\calH$.
By the positivity of $T\ten\id_{M_2}$ we have for all $A\in\A$,
   $$\inp{\psi\oplus T(A)\psi}{\bigl(T\ten\id\bigr)
     \left(\begin{bmatrix}A&-\one\cr0&0\end{bmatrix}^*
                       \begin{bmatrix}A&-\one\cr0&0 \end{bmatrix}\right)
                       \psi\oplus T(A)\psi}\ge0\;.$$
Writing this out we obtain
   $$\inp{\psi}{\bigl(T(A^*A)-T(A)^*T(A)\bigr)\psi}\ge0\;.$$
\end{proof}

\begin{cor} \label{C 3.3}
If  $T:(\A,\ph)\to(\B,\psi)$ then for all $A\in\A$
   $$\ph(T(A)^*T(A))\le\ph(A^*A)\;.$$
\end{cor}

This inequality states that $T$ is a contraction between the GNS
Hilbert spaces of $(\A,\ph)$ and $(\B,\psi)$.

\begin{lem} \label{L 3.4}
$T:(\A,\ph)\to(\B,\psi)$ is an isomorphism in the category QP
if and only if $T:\A\to\B$ is a *-isomorphism.
\end{lem}

\begin{proof}
(Exercise:)
Apply Schwartz's inequality to $T$ and to $T^{-1}$.
\end{proof}

\medskip\noindent
A {\it random variable} (cf. Events and random variables,
in Section \ref{one}) is an injective *-homomorphism
    $$j:\Aph\to\Aphh\;.$$
A {\it quantum stochastic process} (\cite{[AFL]}) is a family $(j_t)_{t\in\TT}$
of random variables indexed by time $\TT$.
Here, $\TT$ is a linearly ordered set such as
$\ZZ$, $\RR$, $\NN$ or $\RR_{+}$.
If $\TT=\RR$ or $\RR_{+}$ we require that for all $A\in\A$ the curve
$t\mapsto j_t(A)$ is strongly continuous.

If $\TT$ is a group, say $\ZZ$ or $\RR$,
then the process is called {\it stationary}
provided that $j_t=\Th_t\circ j_0$ for some representation $t\mapsto\Th_t$
of $\TT$ into the automorphisms of $(\Ah,\phh)$.

\subsubsection{Open system interpretation}

We are observing a subsystem with observable algebra $\A$
of a larger environment with algebra $\Ah$ that we cannot see.
In the Heisenberg picture, the smaller algebra is
moving inside the larger one.
If $t_0\le t_1\le\cdots\le t_n$ is a sequence of times,
and $\tuple E_n$ a sequence of events in $\A$, then
   $$\phh
\bigl(\jtE_1\jtE_2\cdots\jtE_{n-1}\jtE_n\jtE_{n-1}\cdots\jtE_2\jtE_1\bigr)$$
is the probability that $E_1$ occurs at time $t_1$, $E_2$ at time $t_2$,
$\ldots$, and $E_n$ at time $t_n$.
Note the double role played here by the time ordering:
Unless some of the questions $\jtE_k$ recur, that is they lie
in $j_t(\A)$ for different values of $t$,
they must be asked in the order dictated by the times $t_k$.

\subsubsection{Stochastic process interpretation}
In a classical stochastic process $(X_t)_{t\in\TT}$
the random variable $X_t$ is a different one for different times $t$,
so the events concerning $X_t$ change in time accordingly.
If the process is stationary, $X_t$ and $X_s$ differ by an automorphism
of the underlying probability space.
These observations generalise to the noncommutative situation.

\subsection{Conditional expectations and transition operators}
If we are to describe an open quantum system such as the damped harmonic
oscillator by an internal dynamics, say $T_t:\A\to\A$,
without reference to its surroundings,
we need to be able to keep track of an observable $A$ which starts in $\A$
at time zero,
during its motion away from the algebra $\A$ at positive times.
That is, we need its conditional expectation.

In view of the discussion of conditional expectations in finite
dimensions, we give the following general definition.

\begin{defn}
Let $j:(\A,\ph)\to(\Ah,\phh)$ be a random variable.
The {\it conditional expectation} (if it exists) is the unique
morphism $P:(\Ah,\phh)\to(\A,\ph)$ for which
   $$P\circ j=\id_\A\;.$$

\end{defn}

Without proof we state some properties.

\begin{propn} \label{P 3.5}
If $P:(\Ah,\phh)\to(\A,\ph)$ is the conditional expectation with respect
to $j:(\A,\ph)\to(\Ah,\phh)$,
then
   $$\forall_{B_1,B_2\in\A}\forall_{A\in\Ah}:B_1P(A)B_2=P(j(B_1)Aj(B_2))\;.$$
In particular
   $$\forall_{F\in\E(\A)}\forall_{A\in\A}:\phh(j(F)Aj(F))=\ph(FP(A)F)\;.$$
\end{propn}

The second line indicates the connection with Theorem \ref{T 3.1}.

\subsection{Markov processes}
Let us now apply the above notion to an open quantum system.

\subsubsection{Two-time-probabilities}
Suppose that for all $s\in\TT$ there exists a conditional expectation $P_s$
with respect to $j_s$.
Then the probability for $F$ to occur at time s and $E$ at time $t\geq s$
can be written as
   $$\phh(j_s(F)j_t(E)j_s(F))=\ph(FP_s(j_t(E))F)
                             =\ph(FT_{s,t}(E)F)\;,$$
where $T_{s,t}=P_s\circ j_t$ is an operation on $(\A,\ph)$,
the {\it transition operator} from time $s$ to time $t$.

\subsubsection{Multi-time-probabilities}
This reduction to the subsystem succeeds for more than two time
points if there also exist conditional expectations $P_{(-\infty,t]}$
onto the algebras
   $$\A_{(-\infty,t]}:=\vN{j_s(\A)}{s\le t}\;.$$
and moreover the {\it Markov property} holds:
\begin{equation} \label{Markov}
   t\le s\Implies P_{(-\infty,t]}(j_s(\A))\subset j_t(\A)\;.
\end{equation}

\begin{propn} \label{P 3.6}
Let $\bigl(j_t:(\A,\ph)\to(\Ah,\phh)\bigr)_{t\in\TT}$
be a Markov process with conditional expectations $P_t$.
Then the transition operators form a monoid:
   $$0\le s\le t\le u\Implies T_{s,t}T_{t,u}=T_{s,u}\;.$$
In particular, if the process is stationary,
then $T_t:=T_{0,t}=T_{s,s+t}$ satisfies
   $$T_sT_t=T_{s+t}\;\qquad(s,t\ge0).$$
\end{propn}

In the latter case, $(T_t)_{t\ge0}$ is known as the
{\it dynamical semigroup} induced by the stationary Markov process.
Conversely, the process $(j_t)_{t\in\TT}$ is called a {\it Markov dilation}
(\cite{[Kum]}) of the dynamical semigroup $(T_t)_{t\in\TT}$.

The situation is symbolised by the commutative diagram
\begin{equation} \label{dilationgen}
   \xymatrix{
{(\A , \varphi)} \ar[r]^{T_t}  \ar[d]_{j} & {(\A , \varphi)}  \\
{(\wh{\A}, \wh{\varphi})}\ar[r]^{\widehat{T_t}}&{(\wh{\A}, \wh{\varphi})} \ar[u]_{P}      }
\end{equation}

Our goal is to describe a Markov dilation of the damped harmonic oscillator.


\section{Second quantisation}  \label{four}
A quantum model of $n$ harmonic oscillators is obtained by taking
the $n$-fold tensor product of the representation
$z\mapsto W(z):=\exp\bigl(i(\Im z)Q-i(\Re z)P\bigr)$ of the
canonical commutation relation (CCR) over $\CC$.
This turns out to be equivalent to a single representation of the CCR
over $\CC^n$.
An infinity of harmonic oscillators
is obtained by replacing $\CC$ with an infinite dimensional separable
Hilbert space $\K$.
It depends on the spectrum of the time evolution on $\K$
(discrete or continuous),
whether a countable infinity of oscillators is obtained or a continuum,
that is a quantum field.
In our dilation of the damped harmonic oscillator we shall need a quantum
field.

As in the case of a single oscillator we have the choice between
different concrete representations:
we may emphasise the field aspect of the construction,
like in the Gaussian representation of the harmonic oscillator,
or the particle aspect of it, like in its matrix representation.
(The Schr\"odinger representation on $L^2(\RR)$ as in
\eqref{fivea}, \eqref{nine}
has no analogue in infinite dimension,
since there exists no Lebesgue measure on $\RR^\infty$.)

The following definition generalises the Weyl relation
\eqref{WeylC} over $\CC$ to that over
a general complex Hilbert space $\K$.
If $\K$ is the $L^2$-space of some measure space $(X,\mu)$,
then $\K$ may be considered as the `quantisation' of $X$,
and the construction below as its `second quantisation'.

We refer to Mark Fannes' lectures in these volumes.

\subsection{The functor $\G$}

\begin{defn}
Let $\K$ be a complex Hilbert space.
A {\it representation of the Canonical Commutation Relations
\tu{(}CCR\tu{)} over $\K$} is a map
$W$ from $\K$ to the unitary operators on some Hilbert space $\calH$
such that for all $f,g\in\K$:
\begin{equation} \label{WeylK}
   W(f)W(g)=e^{-i\Im\inp fg}W(f+g)\;,
   \end{equation}
 and $t \mapsto W(tf) \psi $ is continuous for all
 $f \in \K , \psi \in \calH$.
The map is called a {\it vacuum representation} if there is a unit
vector $\O\in\calH$ such that
   $$\inp{\O}{W(f)\O}=e^{-\hf\norms f}\;.$$
A vacuum representation is called {\it cyclic} if
the linear span of the vectors $W(f)\O$ is dense in $\calH$.
\end{defn}

A cyclic vacuum representation of the CCR over $\K$ can be constructed
by a generalisation of the method used several times in
the harmonic oscillator in Section \ref{two}:
Let $\pi$ be a minimal Kolmogorov decomposition of the positive definite kernel
$\K\times\K\to\CC$ mapping $(f,g)$ to $e^{\inp fg}$,
and on the total set of `coherent vectors' $\pi(g)$, define
   $$W(f)\pi(g):=e^{-\inp fg-{1\over2}\norms f}\pi(f+g)\;,
               \quad(f,g\in\K)\;.$$
Then put $\O:=\pi(0)$,
and all the requirements in the above definition are met.
On the other hand,
given any cyclic vacuum representation of the CCR over $\K$ with vacuum
vector $\O$,
a Kolmogorov decomposition $\pi'$ of the above
mentioned kernel is obtained by putting
   $$\pi':f\mapsto e^{{1\over2}\norms f}W(f)\O\;.$$
Indeed,
\begin{align*}
  \inp{\pi'(f)}{\pi'(g)}
        &=e^{{1\over2}(\norms f+\norms g)}\cdot e^{i\Im\inp fg}
            \inp\O{W(-f+g)\O}\cr
        &=e^{{1\over2}(\norms f+\norms g)}\cdot e^{i\Im\inp fg}
          e^{-{1\over2}\norms{f-g}}\cr
        &=e^{\inp fg}\;.
\end{align*}
Thus all cyclic vacuum representations of the CCR over a Hilbert space $\K$ are
unitarily equivalent.
However,
this can not be concluded from von Neumann's uniqueness theorem,
since the latter breaks down for infinite dimesional $\K$.
In this case there are indeed many inequivalent (non-vacuum) representations,
for instance those associated to positive temperatures (\cite{[BrR]}).

Since
$e^{\inp{f_1\oplus f_2}{g_1\oplus g_2}}
=e^{\inp{f_1}{g_1}}\cdot e^{\inp{f_2}{g_2}}$,
a representation of the CCR over a direct sum $\K_1\oplus\K_2$
of Hilbert spaces is isomorphic to the tensor product of the representations
of the CCR over $\K_1$ and $\K_2$.

\begin{defn}
Let $\W_0(\K)$ denote the linear span of the
operators $W(f)$, $f\in\K$ in some representation of the CCR over $\K$.
Let $\W(\K)$ be its strong closure.
On the von Neumann algebra $\G(\K)$ we assume by default the vacuum state
   $$\ph_\K(W(f))=e^{-\hf\norms f}\;.$$
Thus $\W(\K)=(\W(\K),\ph_\K)$ is a quantum probability space.

\end{defn}

If $C$ is a contraction $\K_1\to\K_2$,
let $\W_0(C):\W_0(\K_1)\to\W_0(\K_2)$ be given by
\begin{equation} \label{defW}
   \W_0(C)\bigl(W(f)\bigr):=e^{\hf(\norms{Cf}-\norms f)}W(Cf)\;.
   \end{equation}

\begin{propn} \label{P 4.1}
The operator $\W_0(C)$ has a unique strongly continuous extension to an
operation $\W(C)$ of $\W(\K)$.
\end{propn}

\begin{proof}
Cf. for instance \cite{[Tee]}.\end{proof}

\begin{rem}
Second quantisation is a functor $\W$ from the category of Hilbert spaces
with contractions to the category of quantum probability spaces
with operations.

\end{rem}

\subsection{Fields}
From the Weyl relation \eqref{WeylK} it follows that $\l\mapsto W(\l f)$
is a strongly continuous unitary representation of $\RR$.
By the spectral theorem there exists a self-adjoint operator $\Phi(f)$
on $\G(\K)$ such that
   $$W(\l f)=e^{i\l\Phi(f)}\;.$$
The Weyl relations then imply that
   $$[\Phi(f),\Phi(g)]=2i\Im\inp fg\cdot\one\;,$$
and in the vacuum state $\ph_\K$ the random variable $\Phi(f)$ has
normal distribution with mean 0 and variance $\norms f$.
The random variables $\Phi(f)$ and $\Phi(g)$ are compatible
if the inner product $\inp fg$ is real, and independent if it is zero.

In particular,
if $\K=L^2(\RR,2\eta dx)$ (as we shall need in Sections \ref{five}
 and \ref{six}),
then by putting
\begin{equation} \label{Brown}
B_t:=\begin{cases}\Phi(1_{[0,t]})&\text{if }t\ge0,\\
                -\Phi(1_{[t,0]})&\text{if }t<0, \end{cases}
\end{equation}
a stochastic process $(B_t)_{t\in\RR}$ is defined with compatible
normally distributed independent increments having variance
   $$\ph_\K\bigl((B_t-B_s)^2\bigl)=|t-s|\;.$$
Thus $B_t$ is a classical Brownian motion.

\subsection{Particles}
A natural choice for the representation space of the CCR over $\K$
is the {\it Fock space} $\F(\K)$ from Example (d) at the end of
 Section \ref{one}:
   $$\F(\K):=\bigoplus_{n=0}^\infty {1\over{n!}}\K^{\ten_{\rm symm}n}\;,$$
with $\pi(f)$ given by the exponential vectors.
Given a contraction $C:\K\to\K$ let $\F(C)$ be the contraction
$\F(\K)\to\F(\K)$ mapping $\pi(f)$ to $\pi(Cf)$ for every $f\in\K$.
This map can be written
   $$\F(U)=\bigoplus_{n=0}^\infty C\ten C\ten\cdots\ten C\;.$$
Given an orthogonal projection $P$ on $\K$
let a self-adjoint operator $d\F(P)$ on $\F(\K)$ be defined by
   $$e^{i\l d\F(P)}:=\F\left(e^{i\l P}\right)\;.$$
This operator $d\F(P)$ is interpreted as the random variable that
counts for how many particles the `question' $P$ is anwered `yes'.
In particular the total number of particles $N$ equals $d\F(\one)$.

If $\K=L^2(\RR,2\eta dx)$,
the Fock space $\F(\K)$ can be written as $L^2(\Delta(\RR),\mu_\eta)$
where $\Delta(\RR)$ is Guichardet's space over $\RR$ (\cite{[Gui]},
\cite{[Maa]} --- see also the lectures by Martin Lindsay in these
volumes).
   $$\Delta(\RR):=\set{\s\subset\RR}{\#(\s)<\infty};$$
and $\mu_\eta$ is the measure on $\Delta$ given by
   $$\mu_\eta(\{\emptyset\})=1\;,$$
   $$\mu_\eta(d\s)=(2\eta)^n\,ds_1ds_2\cdots ds_n
             \text{ if }\s=\{s_1,s_2,\cdots,s_n\}.$$
The coherent vectors are represented as the functions
$\pi(f):\Delta(\RR)\to\CC$ given by
   $$\s\mapsto\prod_{s\in\s} f(s)\;.$$
Indeed,
\begin{align*}
  \inp{\pi(f)}{\pi(g)}
 &=\int_{\Delta(\RR)}\overline{\pi(f)(\s)}\pi(g)(\s)\mu_\eta(d\s)\cr
 &=\szi n(2\eta)^n\int_{s_1\le\cdots\le s_n}(\bar fg)(s_1)\cdots(\bar fg)(s_n)
   ds_1\cdots ds_n\cr
 &=\szi n{{(2\eta)^n}\over{n!}}\int_{\RR^n}(\bar fg)(s_1)\cdots(\bar fg)(s_n)
   ds_1\cdots ds_n\cr
 &=\szi n{1\over{n!}}\left(2\eta\intR\bar{f(s)}g(s)ds\right)^n
  =e^{\inp fg}\;.
  \end{align*}

In this concrete representation the number operator $d\F(P)$,
where $P$ is the multiplication in $L^2(\RR)$ by $1_S$,
counting the number of particles in the region
$S\subset\RR$, is itself a multiplication operator,
multiplying by the number $\#(\s\cap S)$ of quanta in $S$.
This is seen by the following calculation:
for all $\l\in\RR$, $f\in\K$ and $\s\in\Delta(\RR)$,
   \begin{align*}
   \left(e^{i\l d\F(P)}\pi(f)\right)(\s)
             &=\pi\left(e^{i\l P}f\right)(\s)\cr
             &=\prod_{t\in\s}e^{i\l 1_S(t)}f(t)\cr
        & =\exp\left(i\l\sum_{t\in\s}1_S(t)\right)\cdot\pi(f)(\s)
             =e^{i\l\#(\s\cap S)}\cdot\pi(f)(\s)\;.
             \end{align*}

\section{Unitary dilations of spiraling motion}  \label{five}
In preparation for the solution of the physical problem of damping
posed at the end of Section \ref{two},
we now consider embeddings of the spiraling evolution \eqref{damposc}
into a unitary one.
Let us describe the spiral by
 \begin{equation} \label{spiral}
   C_t:\CC\to\CC:z\mapsto e^{(-\eta+i\o)t}z\;,\qquad(t\ge0).
\end{equation}

\begin{thm}
[{\rm(Sz.~Nagy, Foias 1953; special case.}] \label{T 5.1}
Up to unitary equivalence there exists a unique Hilbert space $\K$
with a unit vector $v$ and a one-parameter group of unitary transformations
$U_t$ on $\K$ such that the span of the vectors $U_tv$, $t\in\RR$ is
dense in $\K$ and
   $$\inp{v}{U_tv}=e^{(-\eta+i\o)t}\;,\quad(t\ge0).$$
\end{thm}

\begin{proof}
{\sl Existence:}
Take $\K:=L^2(\RR,2\eta\,dx)$, let $U_t$ be the shift to the right, and
   $$v(x):=\begin{cases}0& \text{if }x>0,\cr
    e^{(\eta-i\o)t}&\text{if }x\le0.\end{cases}$$
Then we arrive at Example \ref{ex 1.7} (e) at the end of Section \ref{one}.
Uniqueness follows from Theorem \ref{T 1.6}.
\end{proof}

The structure $(\K,J,U_t)$, illustrated in the diagram below,
is called a {\it minimal unitary dilation} of $(C_t)_{t\ge0}$.
\begin{equation} \label{dilationlin}
   \xymatrix{
{\quad \CC \quad} \ar[r]^{e^{(-\eta +i\omega )t}}  \ar[d]_{J:z \mapsto zv} & {\CC}  \\
{\K}\ar[r]^{U_t}&{\K} \ar[u]_{J^\ast : k \mapsto \la v,k \ra}      }
\end{equation}
In practice several --- unitarily equivalent ---
minimal unitary dilations of $(C_t)_{t\ge0}$ can be useful.
If $\K=L^2(\RR)$ and $U_t$ is the shift, then
we speak of {\it translation dilations} of $(C_t)_{t\ge0}$.
They differ only in the shape of $v\in L^2(\RR)$,
which must satisfy
   $$|\hat v(\l)|^2={1\over{(\l-\o)^2+\eta^2}}\;,\quad(\l\in\RR)\;.$$
Particular solutions are $\hat v_\pm(\l):=1/(\l-\o\pm i\eta)$.
Here $v_{+}$,
which occurred in the proof of Theorem \ref{T 5.1},
leads to the {\it incoming translation dilation}
and $v_{-}$ to the {\it outgoing translation dilation}:
   $$v_{-}(x)=\begin{cases}e^{-(\eta+i\o)x}&(x\ge0);\cr 0&(x<0).\end{cases}$$
The former is more useful for the study of incoming fields and particles,
the latter for outgoing ones.
We shall have occasion to employ both below.
The unitary equivalence of these two unitary dilations,
asserted by Theorem \ref{T 5.1},
is implemented by the {\it scattering operator} $S$ which in terms of the
Fourier transform $F$ can be written as
   $$S:=F M_s F^{-1}\;,\quad s(\l):={{\l-\o+i\eta}\over{\l-\o-i\eta}}\;.$$
Apart from these two translation dilations,
the {\it interaction dilation} $(\K,J,U_t)$,
where $\K=L^2(-\infty,0]\oplus\CC\oplus L^2[0,\infty)$,
$J:z\mapsto 0\oplus z \oplus 0$, and $U_t$ describes a more complicated
coupling of $\CC$ to an {\it incoming} and an {\it outgoing channel},
is physically more enlightening, but too cumbersome to treat here.
We refer to \cite{[KuS]} for a thorough treatment.


\section{The damped harmonic oscillator} \label{six}
Equipped with the notions introduced in Sections \ref{one},
\ref{three}, \ref{four}, and \ref{five} we are
now in a position to answer the questions posed at the end of Section \ref{two}.

We act with the second quantisation functor $\W$ of Section \ref{four} on all four
corners and all four arrows of the dilation diagram \eqref{dilationlin}
of Section \ref{five}.
The corners become quantum probability spaces (Section \ref{one}),
and the arrows become operations (Section \ref{three}):
\begin{equation} \label{dilationfunc}
   \xymatrix{
{\G(\CC)} \ar[r]^{\G (C_t)}  \ar[d]_{\G(J)} & {\G(\CC)}  \\
{\G(\K)}\ar[r]^{\G(U_t)}&{\G(\K)} \ar[u]_{\G(J^\ast)}      }
\end{equation}
The answers to Questions 1 and 2 can now be read off.

\begin{enumerate}
\item[(1)]
$\W(\CC)=\B(\calH)$,
where $\calH=\calH_H=l^2(\NN,{1\over{n!}})$ or equivalently
$\calH=\calH_G=L^2(\RR,\g)$ is the Hilbert space of the harmonic oscillator in the
Heisenberg or in the Gaussian representation.
The damped time evolution $T_t:\B(\calH)\to\B(\calH)$ is now given by
   $$T_t:=\W(C_t)=\W\left(e^{(-\eta+i\o)t}\right)\;.$$
Then
\begin{equation} \label{formTt}
  T_t(W(z))=e^{\hf(e^{-2\eta t}-1)|z|^2}W(e^{(-\eta+i\o)t}z)\;.
  \end{equation}
By substituting $W(t+is)=e^{-itP+isQ}$ and differentiating with respect
to $t$ and $s$ respectively, we indeed obtain the equations
\eqref{damposc}.

\item[(2)]
The diagram shows how $T_t$ is embedded into a larger whole,
where the time evolution is a one-parameter group of *-automorphisms,
that is reversible.
Here $j:=\W(J)$ is an injective *-homomorphism,
$P=\W(J^*)$ is a conditional expectation.
By Theorem \ref{T 5.1} this is the only quasifree dilation,
that is in the range of the functor $\W$.
It is automatically Markov.

\end{enumerate}

In this Section we shall discuss four aspects of the construction:
the stochastic behaviour of the oscillator (spirals),
its driving field (a quantum Brownian motion),
the jumps between the levels of the oscillator (a death process),
and the outgoing quanta (a point process).
A complete picture would include the outgoing field and the scattering of
incoming particles as well.
This can easily be achieved using the tools developed here.

\subsection{Stochastic behaviour of the oscillator}
By the functorial character of $\W$ we can split $T_t$ as
   $$T_t=\W(e^{(-\eta+i\o)t})=\W(e^{i\o t})\W(e^{-\eta t})\;.$$
The operator $\W(e^{-i\o t})$ is the automorphism $\a_t$ studied
in Section \ref{three}.
So let us now look at the `dissipative' part $\W(e^{-\eta t})$.

\begin{propn} \label{P 6.1}
For $0\le c<1$ the operator $\W(c)$ leaves invariant the abelian subalgebras
generated by $\one$ and any of the operators $xP-yQ$ with $(x,y)\in\RR^2$.
In particular its action on the algebra
   $$\Q:=\set{f(Q)}{f\in\Linf(\RR)}$$
is given by
\begin{equation} \label{diffsem}
   \W(c)(f(Q))
     ={1\over\sqrt{2\pi(1-c^2)}}
      \intR e^{-{{x^2}\over{2(1-c^2)}}}\;f(cQ+x)\,dx\;.
      \end{equation}

 \end{propn}
\begin{proof}
Obviously, $\W(c)$ leaves the linear span of $\set{W(\l z)}{\l\in\RR}$
invariant, and thus also its strong closure by Proposition \ref{P 4.1}.
Putting $f(Q)=e^{iyQ}$ the r.h.s. of \eqref{diffsem} equals
   $$\left({1\over\sqrt{2\pi(1-c^2)}}\intR e^{-{{x^2}\over{2(1-c^2)}}}
     e^{ixy}dx\right)\cdot e^{icyQ}
    =e^{-{1\over2}(1-c^2)y^2}W(icy)\;,$$
which is equal to the l.h.s. by the definition \eqref{defW} of $\W$.
The theorem follows from the strong continuity of $\W(c)$.
\end{proof}

We recognise the semigroup $T_t$ of transition operators
restricted to $\Q$ as the transition operators of a diffusion
on $\RR$ with a drift towards the origin proportional to the distance
to the origin.

\subsection{The driving field}  \label{sub 6.2}
Let us consider the dilation of the semigroup $T_t$ for $\o=0$.
We take the second quantised {\it incoming} translation dilation
of Section \ref{five},
and substitute it into the diagram \eqref{dilationfunc}.
Let $B_t$ be the Brownian motion given by \eqref{Brown},
and let $\hat Q_t$ denote the embedded oscillator $\Phi(v_t)$.

\begin{propn} \label{P 6.2}
The embedded oscillator $\hat Q_t$ satisfied the integral equation
\begin{equation} \label{SDEQ}
   \hat Q_t-\hat Q_s=-\eta\int_s^t\hat Q_u du+B_t-B_s\;.
   \end{equation}
\end{propn}

This is the integral version of the stochastic differential equation
   $$d\hat Q_t=-\eta\hat Q_t dt +dB_t\;.$$
So we find an embedded Ornstein-Uhlenbeck process in our Markov
dilation.

\begin{proof}
(\cite{[LeT]})
The following equality between functions in $\K$ holds:
   $$U_tv_{-}-U_sv_{-}=-\eta\int_s^t v_u du+1_{[s,t]}\;,\quad(s\le t)\;.$$
Acting with $\Phi$ on both sides of the equation yields \eqref{SDEQ}.
\end{proof}

\subsection{Quanta}
We now concentrate on another abelian subalgebra of $\W(\CC)$,
namely the algebra of all diagonal matrices in Heisenberg's matrix
mechanics.
In terms of the operator $N$ denoting the number of excitations
of the oscillator, this algebra can be written as
   $$\N:=\set{f(N)}{f\in l^\infty(\NN)}\sim l^\infty(\NN)\;.$$

This time we need not put $\o=0$.
Let $\ph_n$ denote the state $l^\infty(\NN)\to\CC:f\mapsto f(n)$.

\begin{propn} \label{P 6.2a}
The diagonal algebra $\N$ is invariant for $T_t$ and
   $$\ph_n\left(T_t\left(s^N\right)\right)=\bigl(1-e^{-2\eta
t}(1-s)\bigr)^n\;.$$
\end{propn}

This is the probability generating function
of a pure death process \cite{[GrS]} with the generator
   $$L=2\eta\begin{bmatrix}0&0&0&0&\cdots\cr
                  1&-1&0&0&\cdots\cr
                  0&2&-2&0&\cdots\cr
                  0&0&3&-3&\cdots\cr
                  \vdots&\vdots&\vdots&\vdots&\ddots\end{bmatrix}\;.$$

\begin{proof} (Sketch. Cf. \cite{[Tee]} for the detailed proof.)
We can write $s^N$ as a weak integral over the operators $W(z)$:
\begin{equation} \label{sNint}
   s^N={1\over\pi(1-s)}\int_\CC e^{-\hf{{1+s}\over{1-s}}|z|^2}W(z)\l(dz)\;,
   \end{equation}
where $\l$ denotes the two-dimensional Lebesgue measure on $\CC$.
This relation can be checked by taking matrix elements
with respect to coherent vectors.
Application of $T_t$ to both sides of \eqref{sNint} yields for all $u,v\in\CC$,
   $$\inp{\pi(u)}{T_t\left(s^N\right)\pi(v)}
              =e^{\bar uv(1-e^{-2\eta t}(1-s))}\;.$$
Since this expression is not sensitive to the relative phase of $u$ and $v$,
the operator $T_t\left(s^N\right)$ lies in $\N$.
The statement is proved by comparing the coefficients of $(\bar uv)^n$ on
both sides.
\end{proof}

\subsection{Emitted quanta}
Finally,
let us see what happens outside the oscillator while
it is cascading down its energy spectrum.
Since we are interested in outgoing quanta at positive times,
let us now consider the {\it outgoing} translation dilation
of $(T_t)_{t\ge0}$ and represent $\W(\K)$ on the Fock space
$L^2(\Delta(\RR))$.
We denote the number operator $d\F(P_{v_t})$ counting the excitations of the
oscillator at time $t$ by $N_t$.

From Proposition \ref{P 6.2a} it follows that the diagram
\eqref{dilationfunc}
can be restricted to the subalgebra $\N\sim l^\infty(\NN)$
\[
 \xymatrix{
{\N} \ar[r]^{T_t}  \ar[d]_{j} & {\N}  \\
{\G(\K)}\ar[r]^{\G(U_t)}&{\G(\K)} \ar[u]_{P}      }
\]
Here
\begin{align*}
  j:&=\G(J):f\mapsto f(N_0)\;;\cr
              P:&=\G(J^*):X\mapsto
 \bigl(\inp{v^{\ten n}}{Xv^{\ten n}}\bigr)_{n=0}^\infty\in l^\infty(\NN).
 \end{align*}
However,
since for different times $t$ and $s$ the functions $v_t$ and $v_s$ are neither
parallel nor orthogonal,
the one-dimensional projections $P_{v_t}$ and $P_{v_s}$ do not commute.
And since for $\l,\mu\in\RR$
   $$e^{i\l N_t}\cdot e^{i\mu N_s}
          =\F\left(e^{i\l P_{v_t}}\right)\F\left(e^{i\mu P_{v_s}}\right)
          =\F\left(e^{i\l P_{v_t}}e^{i\mu P_{v_s}}\right)\;,$$
the number operators $N_t$ and $N_s$ do not commute either.
So the embedded algebras $j_t(\N)$ with $t\in\RR$ do not generate
an abelian subalgebra of $\G(\K)$,
as was the case for the algebras $j_t(\Q)$ above.

For every $t\in\RR$ let us consider the following three number operators.
 \begin{align*}
 N_t&:=d\F(P_{v_t}),
       \quad\hbox{the number of quanta in the oscillator,}\cr
              M_t&:=d\F\left(M_{1_{[t,\infty)}}\right)
 \quad\hbox{the number of quanta that have not yet left the oscillator,}\cr
              K_t&:=d\F\left(M_{1_{(-\infty,t]}}\right)
 \quad\hbox{the number of outgoing quanta that have left the oscillator.}
 \end{align*}
Note that the number $M_t-N_t$ of incoming quanta is not given by a
multiplication operator, but the number $K_t$ of outgoing quanta is.
This is due to the fact that we are considering the outgoing
translation dilation of $T_t$.
Note furthermore that the operators $M_t$ and
$K_s$ ($s,t \in \RR$) all commute. 

For positive times the number $M_t-N_t$ of incoming quanta
has expectation 0 in the states of the form $\th\circ P$
($\th\in l^1(\NN)$) which we consider.
So we may expect that replacing $N_t$ by $M_t$ would lead to an
embedded classical Markov chain.

\begin{propn} \label{P 6.3}
For $t\ge0$ we have the following commuting diagram involving abelian
von Neumann algebras.
\[
 \xymatrix{
{\N} \ar[r]^{T_t}  \ar[d]_{j:f\mapsto f(M_0)} & {\N}  \\
{L^\infty (\Delta ,\mu_\eta)}\ar[r]^{\G(U_t)}&{L^\infty (\Delta ,\mu_\eta)}
\ar[u]_{P=\G(J^\ast)}      }
\]

\end{propn}

\begin{proof}
For all $z,u\in\CC$, $t\ge0$ and $s\in[0,1]$ we have
\begin{align*}
\inp{\pi(u)}{P\circ\G(U_t)\circ j(s^N)\pi(z)}
    &=\inp{\pi(u)}{P\left(s^{M_t}\right)\pi(z)}\cr
    &=\szi n(\bar uz)^n\int_{\Delta_n(\RR)}s^{M_t(\s)}|v^{\ten n}(\s)|^2
          \mu_\eta(d\s)\cr
    &=\szi n(2\eta\bar uz)^n\cdot{1\over{n!}}\int_{\RR^n}
          \left(\prod_{j=1}^n s^{1_{[t,\infty)}(r_j)}|v(r_j)|^2\right)
          dr_1\cdots dr_n\cr
    &=\exp\left(2\eta\bar uz\int_0^\infty
          e^{-2\eta r}s^{1_{[t,\infty)}(r)}dr\right)\cr
    &=\exp\left(\bar uz
   \left(-e^{-2\eta r}\from0to t-se^{-2\eta r}\from t to\infty\right)\right)\cr
    &=\exp\left(\bar uz(1-e^{-2\eta t}(1-s))\right)\;.
    \end{align*}
By Proposition \ref{P 6.2}
 the latter expression is equal to $\inp{\pi(u)}{T_t(s^N)\pi(z)}$.
\end{proof}

Finally, since $K_t+M_t$ is equal to the total number of quanta,
which is the same constant for all times,
we conclude that a quantum is emitted at precisely the moment that
the oscillator makes a downward jump.
Moreover, these jumps are made at independent
exponentially distributed random times.

These phenomena turn out to be natural consequences of damped harmonic
motion in a noncommutative description.

\end{document}